\documentclass[a4paper,12pt]{article}
\pdfoutput=1
\usepackage{geometry}
\usepackage{amsmath,amssymb,amsfonts}
\usepackage{xcolor,graphicx,cite,soul}
\usepackage{array,booktabs,longtable}
\usepackage{caption,microtype}
\usepackage{subcaption}
\usepackage{hyperref}
\usepackage{datetime}
\usepackage{appendix}
\usepackage{url}
\usepackage{ulem}

\newcommand{\lsim}
{\;\raisebox{-.3em}{$\stackrel{\displaystyle <}{\sim}$}\;}
\newcommand{\gsim}
{\;\raisebox{-.3em}{$\stackrel{\displaystyle >}{\sim}$}\;}

\newcommand\Code[1]{\ensuremath{\texttt{#1}}}

\newcommand\al{\alpha}
\newcommand\be{\beta}
\newcommand\tb{\tan\beta}

\newcommand\LP{\left(}
\newcommand\RP{\right)}

\newcommand\ReDiag{\mathop{%
  \raise .5pt\hbox{[}%
  \widetilde{\mathrm{Re}}%
  \raise .5pt\hbox{]}}}
\newcommand\ReOffDiag{\mathop{%
  \raise .5pt\hbox{$\llbracket$}%
  \widetilde{\mathrm{Re}}%
  \raise .5pt\hbox{$\rrbracket$}}}

\newcommand\cL{{\cal L}}

\newcommand\MH{m_H}

\newcommand\MHp{m_{H^\pm}}

\newcommand\refeq[1]{Eq.~(\ref{#1})}
\newcommand\refeqs[1]{Eqs.~(\ref{#1})}
\newcommand\refta[1]{Tab.~\ref{#1}}
\newcommand\reftas[1]{Tabs.~\ref{#1}}
\newcommand\refse[1]{Sect.~\ref{#1}}

\newcommand\citere[1]{Ref.~\cite{#1}}
\newcommand\citeres[1]{Refs.~\cite{#1}}

\newcommand{\CP}{{\cal CP}}
\newcommand{\cp}{{\CP}}

\newcommand{\gev}{\,\, \mathrm{GeV}}

\newcommand\fb{\ensuremath{\mbox{fb}}}
\newcommand\ab{\ensuremath{\mbox{ab}}}

\newcommand\iab{\ensuremath{\ab^{-1}}}

\newcommand{\br}{\text{BR}}
\newcommand{\De}{\Delta}

\newcommand{\sig}{\sigma}

\def\reffi#1{\mbox{Fig.~\ref{#1}}}
\def\reffis#1{\mbox{Figs.~\ref{#1}}}

\def\Ga{\Gamma}
\def\ga{\gamma}

\def\la{\lambda}

\def\sgn{\mathrm{sgn}}

\newcommand{\zet}[1]{\ensuremath{\mathbb{Z}_{#1}}}

\definecolor{Orange}{named}{orange}
\definecolor{Purple}{named}{purple}
\definecolor{Lightblue}{cmyk}{0.9,0.1,0.1,0.3}
\definecolor{dgelborange}{cmyk}{0.,0.3,0.5, 0.}
\definecolor{Lila}{rgb}{0.5,0.,1}

\graphicspath{{figs/}}
\renewcommand{\arraystretch}{1.2}

\allowdisplaybreaks
\newcommand{\ns}{\ensuremath{N_S}}
\newcommand{\nb}{\ensuremath{N_B}}
\newcommand{\nt}{\ensuremath{N_T}}
\newcommand{\Ns}[1]{\ensuremath{N_{S,#1}}}

\newcommand{\esb}{\ensuremath{\epsilon_{{\rm syst,}B}}}
\newcommand{\gZ}{\ensuremath{g_Z}}
\newcommand{\gx}{\ensuremath{g_x}}

\newcommand{\non}{\nonumber}
\newcommand{\Gatot}{\ensuremath{\Ga_{\rm tot}}}

\begin{document}
\thispagestyle{empty}

\def\thefootnote{\fnsymbol{footnote}}

\begin{flushright}
\mbox{}
DESY 21-230\\
IFT--UAM/CSIC-21-158\\
\end{flushright}

\mbox{}
\vspace{0.5cm}

\begin{center}

{\large\sc 
  {\bf A 96 GeV Higgs Boson in the 2HDM plus Singlet}}\\

\vspace{1cm}

{\sc
S.~Heinemeyer$^{1}$%
\footnote{email: Sven.Heinemeyer@cern.ch}%
, C.~Li$^2$%
\footnote{email: cheng.li@desy.de}%
, F.~Lika$^3$%
\footnote{email: florian.lika@desy.de}%
, G.~Moortgat-Pick$^{2,3}$%
\footnote{email: gudrid.moortgat-pick@desy.de}%
~and S.~Paasch$^2$%
\footnote{email: steven.paasch@desy.de}
}

\vspace*{.7cm}

{\sl
$^1$Instituto de F\'isica Te\'orica (UAM/CSIC), 
Universidad Aut\'onoma de Madrid, \\ 
Cantoblanco, 28049, Madrid, Spain

\vspace{0.1cm}

$^2$DESY, Notkestra\ss e 85, 22607 Hamburg, Germany

\vspace{0.1cm}

$^3$II. Institut f\"ur Theoretische Physik, Universit\"at Hamburg,\\
Luruper Chaussee 149, 22761 Hamburg, Germany

}

\end{center}

\vspace*{0.1cm}

\begin{abstract}
\noindent
We discuss a $\sim 3\,\sigma$
signal (local) in the light Higgs-boson search
in the diphoton decay mode at $\sim 96 \gev$ as reported by
CMS, together with a $\sim 2\,\sigma$ excess (local)
in the $b \bar b$ final state
at LEP in the same mass range.
We interpret this possible signal as a Higgs boson in the 
2~Higgs Doublet Model type~II with an additional Higgs singlet, which can be either complex (2HDMS) or real (N2HDM).
We find that the lightest $\cp$-even Higgs boson of the two models can
equally yield a perfect fit to both excesses simultaneously, while the second lightest
state is in full agreement with the Higgs-boson measurements at
$125 \gev$, and the full Higgs-boson
sector is in agreement with all Higgs exclusion
bounds from LEP, the Tevatron and the LHC as well as other theoretical and
experimental constraints. 
We derive bounds on the 2HDMS and N2HDM Higgs sectors from a fit to both excesses and describe how this signal can be further analyzed at
future $e^+e^-$~colliders, such as the ILC. We analyze in
detail the anticipated precision of the coupling measurements of the
$96 \gev$ Higgs boson at the ILC. We find that these Higgs-boson measurements at the LHC and the ILC cannot distinguish between the two Higgs-sector realizations.
\end{abstract}


\def\thefootnote{\arabic{footnote}}
\setcounter{page}{0}
\setcounter{footnote}{0}


\newpage
\section {Introduction}
\label{sec:intro}

In the year 2012 the ATLAS and CMS collaborations have discovered a new
particle that is -- within theoretical and experimental uncertainties --
consistent with the existence of a Higgs boson as predicted by the Standard
Model (SM) with a mass
of~$\sim 125 \gev$~\cite{Aad:2012tfa,Chatrchyan:2012xdj,Khachatryan:2016vau}.
So far no conclusive signs of physics beyond the~SM (BSM) have been found at
the LHC. However, the measurements of Higgs-boson production and decay rates
at the LHC, which are known experimentally to a precision of roughly
$\sim 10-20\%$, leave ample room for BSM interpretations.
Many BSM models contain extended Higgs-boson sectors.
Consequently, one of the main tasks of the LHC Run~III and beyond is to
determine whether the observed scalar boson forms part of the Higgs-boson
sector of an extended model. Such extended Higgs-boson sectors naturally
possess additional Higgs bosons, which can have masses larger, but also
smaller than $125 \gev$. Some examples for the latter can be found in
\citeres{Bernon:2014nxa,Robens:2015gla,Heinemeyer:2011aa,Domingo:2015eea,Bechtle:2016kui}
(see also the discussion below). Therefore, 
the search for lighter Higgs bosons forms an
important part in the BSM Higgs-boson program at the (HL-)LHC and planned lepton collider experiments.

Searches for Higgs bosons below $125 \gev$ have been performed at
LEP~\cite{Abbiendi:2002qp,Barate:2003sz,Schael:2006cr},
the Tevatron~\cite{Group:2012zca} and the
LHC~\cite{CMS:2017yta,Sirunyan:2018aui,Sirunyan:2018zut,ATLAS:2018xad}.
LEP observed a $2.3\,\sigma$ local excess
observed in the~$e^+e^-\to Z(H\to b\bar{b})$
channel~\cite{Barate:2003sz}, consistent with a
scalar of mass $\sim 98 \gev$. However, due to the $b \bar b$ final state the  
mass resolution is rather coarse.
ATLAS and CMS searched for light Higgs bosons in the diphoton
final state. The CMS Run~II results~\cite{Sirunyan:2018aui} 
show a local excess of $\sim 3\, \sigma$ at
$\sim 96 \gev$, where a similar excess of $2\, \sigma$ had been observed in
Run~I~\cite{CMS:2015ocq} at roughly the same mass.
First Run\,II~results from~ATLAS, on the other hand, 
using~$80$\,fb$^{-1}$ turned out to be weaker than the corresponding CMS
results, see, e.g., Fig.~1 in~\citere{Heinemeyer:2018wzl}.

Since the CMS and the LEP excesses in the searches for light Higgs-bosons
are found effectively at the same mass, the question arises 
whether they have a common origin, i.e.\ which model can accomodate them
simultaneously, while being in agreement with all other Higgs-boson
related limits and measurements. The two excesses have been described in the
following models (see also \cite{Heinemeyer:2018wzl,Heinemeyer:2018jcd,Richard:2020jfd,Biekotter:2020cjs}):
(i) the Next-to-Two Higgs Doublet model,
N2HDM~\cite{Biekotter:2019kde,Biekotter:2019mib,Biekotter:2019drs,Biekotter:2020ahz,Biekotter:2020cjs,Biekotter:2021qbc},
as will be discussed below (additionally simultaneous explanations for possible excesses at $\sim 400 \gev$ are discussed in \citere{Biekotter:2021qbc});
(ii) various realizations of the Next-to-Minimal Supersymmetric SM, 
     NMSSM~\cite{Biekotter:2021qbc,Domingo:2018uim,Choi:2019jts} (see also \citere{Biekotter:2021qbc} for the possible $\sim 400 \gev$ excesses);
(iii) the $\mu$-from-$\nu$ supersymmetric SM ($\mu\nu$SSM) with
      one~\cite{Biekotter:2017xmf} and three
      generations~\cite{Biekotter:2019gtq} of right-handed neutrinos;
(iv) Higgs inflation inspired $\mu$NMSSM~\cite{Hollik:2018yek};
(v) NMSSM with a seesaw extension~\cite{Cao:2019ofo};
(vi) Higgs singlet with additional vector-like
     matter, as well as Two Higgs Doublet Model,
     2HDM type~I~\cite{Fox:2017uwr};
(vii) 2HDM type~I with a moderately-to-strongly
       fermiophobic CP-even Higgs~\cite{Haisch:2017gql};
      (viii) Radion model~\cite{Richard:2017kot};
(ix) Higgs associated with the breakdown of an $U(1)_{L_\mu L_\tau}$
    symmetry~\cite{Liu:2018xsw};
(x) Minimal dilaton model~\cite{Liu:2018ryo};
(xi) Composite framework containing a pseudo-Nambu Goldstone-type light
     scalar~\cite{Richard:2020jfd}.
(xii) SM extended by a complex singlet scalar field (which can also
      accommodate a pseudo-Nambu Goldstone dark matter)~\cite{Cline:2019okt};
(xiii) Anomaly-free $U(1)'$ extensions of SM with two complex scalar
    singlets~\cite{AguilarSaavedra:2020wmg}. In the model realizations (xii)
    and (xiii) the required di-photon decay rate is reached by adding
    additional charged particles that couple to the 96~GeV scalar.
On the other hand, in the MSSM the CMS excess cannot be
realized~\cite{Bechtle:2016kui}.
The important question arises, how one can distinguish the various model
realizations, in particular when they are similar to each other, e.g.\ the
various supersymmetric (SUSY) realizations.     

In this paper we focus on pure Higgs-sector extensions of the SM, based on the
2HDM type~II (which naturally lead the way to SUSY realizations). As listed above, the 2HDM with an additional real singlet, the N2HDM can describe
simultaeneously the LEP and CMS excesses at
$\sim 96 \gev$~\cite{Biekotter:2019kde}. While this model possesses an
additional real Higgs singlet, it also has an additional $Z_2$~symmetry that
can provide a Dark Matter candidate in case of a zero singlet
vev~\cite{Chen:2013jvg,Muhlleitner:2016mzt,Engeln:2020fld}. Such an additional
$Z_2$ symmetry, however, is not compatible with the Higgs sector of
supersymmetric models, such as the NMSSM (which can equally describe the two
excesses~\cite{Domingo:2018uim,Choi:2019jts}). The ``pure Higgs sector
extension'' of the NMSSM is the 2HDM with an additional complex singlet and an
additional $Z_3$ symmetry, the 2HDMS~\cite{Branco:2011iw,Baum:2018zhf}.
Consequently, to take a first step into model distincion, in this article we analyze for which part of the paramater spaces
the 2HDMS and N2HDM (type~II) can describe the CMS and LEP excesses. We investigate where the models
give the same predictions, where they may differ, and how the two model
realizations could possibly be distinguished from each other.

For the analyses of the 2HDMS and N2HDM Higgs-boson sectors at future colliders we employ
the anticipated reach and precision of the HL-LHC, and furthermore of a possible
future $e^+e^-$ collider, where we focus on the International Linear Collider
(ILC) with a center-of-mass energy of $\sqrt{s} = 250 \gev$ (ILC250).
In particular we show what can be learned from a measurement of the couplings
of the $125 \gev$ Higgs-boson at the ILC250. Going one step further, we also
analyze the coupling measurement of the $96 \gev$ Higgs-boson at the ILC250.

Our paper is organized as follows. 
In \refse{sec:model} we describe the relevant features of the 2HDMS and the
N2HDM, and where they differ from each other.
The experimental and theoretical constraints taken into account are given
in \refse{sec:constraints}.
Details about the experimental excesses at CMS and LEP, as well
as details on their implemenations are discussed 
in \refse{sec:excesses}.
In \refse{sec:results} we present our results in the 2HDMS and contrast them
to the results in the N2HDM. We discuss the possibilities to
investigate these scenarios at the HL-LHC and the ILC250. We conclude
with \refse{sec:conclusions}.

\section{The N2HDM and the 2HDMS}
\label{sec:2hdms}

\subsection{Symmetries and the Higgs potential}
\label{sec:potential}

In this section we define the two models under consideration, the N2HDM and the 2HDMS. Both extend the 2HDM with the doublets $\Phi_1$ and $\Phi_2$ by a singlet $S$, which is taken to be real (complex) for the N2HDM (2HDMS). 
After electroweak symmetry breaking, the scalar component of $\Phi_1$, $\Phi_2$ and $S$ acquire the non trivial vacuum expectation values (vevs). Thus the fields can be expanded around vevs and have the following expressions:
\begin{gather}
	\Phi_1=\begin{pmatrix}
		\chi_1^+\\\phi_1
	\end{pmatrix}=\begin{pmatrix}
		\chi_1^+\\v_1+\displaystyle{\frac{\rho_1+i\eta_1}{\sqrt{2}}}
	\end{pmatrix}
	\Phi_2=\begin{pmatrix}
		\chi_2^+\\\phi_2
	\end{pmatrix}=\begin{pmatrix}
		\chi_2^+\\v_2+\displaystyle{\frac{\rho_2+i\eta_2}{\sqrt{2}}}
	\end{pmatrix}\notag\\
	S=v_S+\frac{\rho_S+i\eta_S}{\sqrt{2}}, 
\end{gather}
where the $\eta_S$ is absent in the N2HDM. 

The models furthermore differ by the symmetry structure. Both models obey the $\zet2$ symmetry that is imposed already on the 2HDM to avoid flavour-changing neutral currents at tree-level (where we will focus on "type~II", see the discussion below). Concerning the singlet in the N2HDM, it is odd under an additional $Z'_2$ symmetry. The 2HDMS, on the other hand, obeys an additional $\zet3$ symmetry. These symmetries can be summarized as follows.

\begin{align}
    {\rm N2HDM~and~2HDMS~}\quad \zet2 &~:~
    \Phi_1 \to \Phi_1, \qquad \Phi_2 \to -\Phi_2, \qquad S\to S \\[.3em]
    {\rm N2HDM~}\quad \zet2' &~:~ 
    \Phi_1 \to \Phi_1, \qquad \Phi_2 \to \Phi_2, \qquad S\to -S \\[.3em]
    {\rm 2HDMS~}\quad \zet3 &~:~ 
    	\begin{pmatrix} \Phi_1\\ \Phi_2\\ S \end{pmatrix} \to 
    	\begin{pmatrix} 
    	1&	&	\\	&	e^{i2\pi/3}&	\\	&	&	e^{-i2\pi/3}
	   \end{pmatrix}\,
	   \begin{pmatrix} \Phi_1\\ \Phi_2\\ S
	\end{pmatrix}
\end{align}

The most general potential of two doublets and one singlet, before applying the additional $Z'_2$ or $\zet3$ symmetry is given by~\cite{Baum:2018zhf} 
\begin{equation}
	\begin{split}
		V=&m_{11}^2\Phi_1^\dagger\Phi_1+m_{22}^2\Phi_2^\dagger\Phi_2-(m_{12}^2\Phi_1^\dagger\Phi_2+\mathrm{h.c.})\\
		&+\frac{\lambda_1}{2}(\Phi_1^\dagger\Phi_1)^2+\frac{\lambda_2}{2}(\Phi_2^\dagger\Phi_2)^2+\lambda_3(\Phi_1^\dagger\Phi_1)(\Phi_2^\dagger\Phi_2)+\lambda_4(\Phi_1^\dagger\Phi_2)(\Phi_2^\dagger\Phi_1)\\
		&+\Big[\frac{\lambda_5}{2}(\Phi_1^\dagger\Phi_2)^2+\lambda_6(\Phi_1^\dagger\Phi_1)(\Phi_1^\dagger\Phi_2)+\lambda_7(\Phi_2^\dagger\Phi_2)(\Phi_1^\dagger\Phi_2)+\mathrm{h.c.}\Big]\\
		&+(\xi S+\mathrm{h.c.})+m_S^2 S^\dagger S+(\frac{{m'_S}^2}{2}S^2+\mathrm{h.c.})\\
		&+\left(\frac{\mu_{S1}}{6}S^3+\frac{\mu_{S2}}{2}S S^\dagger S+\mathrm{h.c.}\right)+\left(\frac{\lambda''_1}{24}S^4+\frac{\lambda''_2}{6}S^2 S^\dagger S+\mathrm{h.c.}\right)+\frac{\lambda''_3}{4}(S^\dagger S)^2\\
		&+\Big[S(\mu_{11}\Phi_1^\dagger\Phi_1+\mu_{22}\Phi_2^\dagger\Phi_2+\mu_{12}\Phi_1^\dagger\Phi_2+\mu_{21}\Phi_2^\dagger\Phi_1)+\mathrm{h.c.}\Big]\\
		&+S^\dagger S\Big[\lambda'_1\Phi_1^\dagger\Phi_1+\lambda'_2\Phi_2^\dagger\Phi_2+\lambda'_3\Phi_1^\dagger\Phi_2+\mathrm{h.c.}\Big]\\
		&+\Big[S^2(\lambda'_4\Phi_1^\dagger\Phi_1+\lambda'_5\Phi_2^\dagger\Phi_2+\lambda'_6\Phi_1^\dagger\Phi_2+\lambda'_7\Phi_2^\dagger\Phi_1)+\mathrm{h.c.}\Big]
	\end{split}
\end{equation}
with 29 free parameters. 

For the N2HDM by applying the $\zet2$ symmetry one finds that
the parameters $\lambda_6,\lambda_7, \lambda'_3, \lambda'_6, \lambda'_7$, which break the $\zet2$ Symmetry explicitly, are zero. Applying the $Z'_2$ symmetry requires all linear or cubic terms to be zero. On the other hand, we keep $m_{12}$, which softly breaks the $\zet2$ symmetry. Furthermore, since the $S$ is real, one has $S^\dagger = S$. Summing up all terms containing $S^2\Phi_1\Phi_1, S^2\Phi_2\Phi_2, S^4$ allows to redefine accordingly the coefficients $\lambda'_1+\lambda'_4$, $\lambda'_2+\lambda'_5$ and $\lambda''_1+\lambda''_2$ a $\lambda_7,\lambda_8, \lambda_6$ respectively (to meet the definitions in~\cite{Muhlleitner:2016mzt}). The potential then reads,
\begin{equation}
\begin{split}
V_\text{N2HDM}=&m_{11}^2\Phi_1^\dagger\Phi_1+m_{22}^2\Phi_2^\dagger\Phi_2-(m_{12}^2\Phi_1^\dagger\Phi_2+\mathrm{h.c.})+\frac{\lambda_1}{2}(\Phi_1^\dagger\Phi_1)^2+\frac{\lambda_2}{2}(\Phi_2^\dagger\Phi_2)^2\\
&+\lambda_3(\Phi_1^\dagger\Phi_1)(\Phi_2^\dagger\Phi_2)+\lambda_4(\Phi_1^\dagger\Phi_2)(\Phi_2^\dagger\Phi_1)+\frac{\lambda_5}{2}[(\Phi_1^\dagger\Phi_2)^2+\mathrm{h.c.}]\\
&+\frac{1}{2}m_S^2 S^2+\frac{\lambda_6}{8}S^4+\frac{\lambda_7}{2}(\Phi_1^\dagger\Phi_1)S^2+\frac{\lambda_8}{2}(\Phi_2^\dagger\Phi_2)S^2~.
\end{split}
\label{eq:n2hdmpot}
\end{equation}

\smallskip
For the 2HDMS by imposing the $\zet3$ Symmetry, we set the $\zet3$ breaking parameters $\lambda_5=\lambda''_1=\lambda''_2=\lambda'_4=\lambda'_5=0$. On the other hand, we keep the terms $m'_S, m_{12}, \mu_{S2}, \mu_{11}, \mu_{22}, \mu_{21}$, which softly break the $\zet3$ symmetry. Taking the mapping of the 2HDMS to the NMSSM in~\cite{Baum:2018zhf} into account, we only keep $m_{12}$ and $\mu_{12}$ as soft breaking parameters. 
The $\zet3$-invariant scalar potential of the 2HDMS, is given by,
\begin{equation}
	\begin{split}
		V_{\rm 2HDMS}=&m_{11}^2(\Phi_1^\dagger\Phi_1)+m_{22}^2(\Phi_2^\dagger\Phi_2)+\frac{\lambda_1}{2}(\Phi_1^\dagger\Phi_1)^2+\frac{\lambda_2}{2}(\Phi_2^\dagger\Phi_2)^2+\lambda_3(\Phi_1^\dagger\Phi_1)(\Phi_2^\dagger\Phi_2)\\
		&+\lambda_4(\Phi_1^\dagger\Phi_2)(\Phi_2^\dagger\Phi_1)+m_S^2( S^\dagger S)+\lambda'_1( S^\dagger S)(\Phi_1^\dagger\Phi_1)+\lambda'_2( S^\dagger S)(\Phi_2^\dagger\Phi_2)\\
		&+\frac{\lambda''_3}{4}( S^\dagger S)^2+\Big(-m_{12}^2\Phi_1^\dagger\Phi_2+\frac{\mu_{S1}}{6} S^3+\mu_{12} S\Phi_1^\dagger\Phi_2+\text{h.c.}\Big)~.
	\end{split}
	\label{eq:2hdmspot}
\end{equation}
In this potential, the parameters $\lambda''_3$, $\lambda'_1$ and $\lambda'_2$ play the similar roles of $\lambda_6$, $\lambda_7$, $\lambda_8$, respectively in the N2HDM Higgs potential shown in \refeq{eq:n2hdmpot}, while the $\lambda_5$ term is forbidden by the $\zet3$ symmetry in the 2HDMS. In addition, we have the new terms $\mu_{12}$ and $\mu_{S1}$ compared to the N2HDM, which arise from the complex singlet.

One can define $\tb := {v_2}/{v_1}$ as in the 2HDM.
Therefore, we obtain the $v = \sqrt{v_1^2+v_2^2} = 174 \gev$ based on our convention of the doublet fields (and again internally for the N2HDM we use a definition with an additional factor of $1/\sqrt{2}$).
By using the minimization conditions:
\begin{equation}
	\frac{\partial V}{\partial \Phi_1}\bigg|_{\substack{\Phi_1=v_1\\\Phi_2=v_2\\S=v_S}}=\frac{\partial V}{\partial \Phi_2}\bigg|_{\substack{\Phi_1=v_1\\\Phi_2=v_2\\S=v_S}}=\frac{\partial V}{\partial S}\bigg|_{\substack{\Phi_1=v_1\\\Phi_2=v_2\\S=v_S}}=0
\end{equation}
$m_{11}^2$, $m_{22}^2$ and $m_{S}^2$ can be replaced by the expression from the tadpole equations. In the 2HDMS we now have 12 free parameters,
\begin{equation}
	\tan\beta,\;\lambda_1,\; \lambda_2,\; \lambda_3,\; \lambda_4,\; \lambda'_1,\; \lambda'_2,\; \lambda''_3,\; m^2_{12},\; \mu_{S1},\; \mu_{12},\; v_S .
	\label{eq:lambdasinp}
\end{equation}
Similarly we have 11 free parameters in the N2HDM,
\begin{equation}
	\tan\beta,\;\lambda_1,\; \lambda_2,\; \lambda_3,\; \lambda_4,\; \lambda_5,\; \lambda_6,\; \lambda_7,\;\lambda_8\; m^2_{12},\; v_S .
	\label{eq:lambdasinp}
\end{equation}


\subsection{Masses and Couplings}
\label{sec:mass-coup}

Due to the configuration of the fields, the charged Higgs sector of the N2HDM and 2HDMS have the same structure as the 2HDM. However, the additional singlet enters the neutral Higgs sector and mixes with two doublets for both CP-even sector and CP-odd sector. This generates an additional scalar Higgs and in the case of the 2HDMS also an additional pseudo-scalar Higgs. In total we have 3 scalar Higgs bosons $h_1$, $h_2$, $h_3$, the charged Higgs boson $H^{\pm}$, as well as 2 pseudo-scalar Higgs bosons $a_1$, $a_2$ for the 2HDMS, or 1 pseudo-scalar Higgs boson $a_1$ for the N2HDM. We apply the conventions of $m_{h_1}<m_{h_2}<m_{h_3}$ and $m_{a_1}<m_{a_2}$ for the remainder of the paper.

We  obtain the CP-even Higgs-boson mass eigenstates by diagonalizing the $3 \times 3$ mass matrix, $M_S^2$. Since this matrix is symmetric, the diagonalization matrix is orthogonal, given by the same $3\times3$ rotation matrix for the 2HDMS and N2HDM. For the CP-even sector, the rotation matrix is reads,
\begin{equation}
	R=\begin{pmatrix}
		c_{\alpha_1}c_{\alpha_2}& s_{\alpha_1}c_{\alpha_2}& s_{\alpha_2}\\
		-s_{\alpha_1}c_{\alpha_3}-c_{\alpha_1}s_{\alpha_2}s_{\alpha_3}& c_{\alpha_1}c_{\alpha_3}-s_{\alpha_1}s_{\alpha_2}s_{\alpha_3}& c_{\alpha_2}s_{\alpha_3}\\
		s_{\alpha_1}s_{\alpha_3}-c_{\alpha_1}s_{\alpha_2}c_{\alpha_3}& -s_{\alpha_1}s_{\alpha_2}c_{\alpha_3}-c_{\alpha_1}{{s_{a}}}_3& c_{\alpha_2}c_{\alpha_3}~,
	\end{pmatrix}
	\label{eq:rot}
\end{equation}
where $\alpha_1$, $\alpha_2$ and $\alpha_3$ are the three mixing angles. The mass basis and the interaction basis are related by,
\begin{equation}
	\begin{pmatrix}
		h_1\\h_2\\h_3
	\end{pmatrix} = R\begin{pmatrix}
		\rho_1\\ \rho_2\\ \rho_S
	\end{pmatrix},\quad \mathrm{diag}\{m^2_{h_1},m^2_{h_2},m^2_{h_3}\}=R^T {M}^2_{S} R~.
\end{equation}

In the CP-odd sector, taking into account the neutral Goldstone boson, in the 2HDMS we need two mixing angles for the diagonalization. The first one is, as in the 2HDM, the angle $\be$. Additionally we define the angle $\alpha_4$ for pseudo-scalar Higgses, and the mixing matrix for CP-odd sector can be expressed as:
\begin{gather}
	R^A=\begin{pmatrix}
		-s_\beta c_{\alpha_4}& c_\beta c_{\alpha_4}& s_{\alpha_4}\\
		s_\beta s_{\alpha_4}& -c_\beta s_{\alpha_4}& c_{\alpha_4}\\
		c_\beta& s_\beta& 0
	\end{pmatrix}
	\label{eq:rota}\\
	\mathrm{with~}	\begin{pmatrix}
		a_1\\a_2\\\xi
	\end{pmatrix}=R^A\begin{pmatrix}
		\eta_1\\ \eta_2\\ \eta_S
	\end{pmatrix},\qquad\mathrm{diag}\{m^2_{a_1},m^2_{a_2},0\}=(R^A)^T {M}^2_{P} R^A~,
\end{gather}
where the $\xi$ denotes the Goldstone boson. 
In the N2HDM this reduces to a $2\times2$ rotation matrix (as in the 2HDM),
\begin{equation}
	R^A=\begin{pmatrix}
        c_{\beta}& s_{\beta} \\
        -s_{\beta}& c_{\beta}~, \\
	\end{pmatrix}
	\label{eq:rotan}
\end{equation}

By using the rotation matrices, one can express the free parameters of the Lagrangian in \eqref{eq:lambdasinp} in terms of the mass of all Higgs bosons and the mixing angles~(see appendix~\ref{appendix:a} for details). As a result we have new sets of input parameters. For the 2HDMS we have
\begin{equation}
	\tan\beta,\quad\alpha_{1,2,3,4},\quad m_{h_1},\quad m_{h_2},\quad m_{h_3},\quad m_{a_1},\quad m_{a_2},\quad m_{H^\pm},\quad v_S
	\label{eq:inpmass}
\end{equation}
and 
\begin{equation}
	\tan\beta,\quad\alpha_{1,2,3},\quad m_{h_1},\quad m_{h_2},\quad m_{h_3},\quad m_{a_1},\quad m^2_{12},\quad m_{H^\pm},\quad v_S~
	\label{eq:inpmassn}
\end{equation}
for the N2HDM. Naturally, the parameters $\al_4$ and $m_{a_2}$ are absent in the N2HDM. However, the different symmetry structure w.r.t.\ the 2HDMS leaves $m_{12}^2$ as additional free parameter. The corresponding situation in the 2HDMS is slightly more involved. When the singlet field acquire the vev, $\mu_{12} S \Phi_1^\dagger\Phi_2$ yields a term $v_S \mu_{12} \Phi_1^\dagger\Phi_2$, which is similar to the term $m_{12}^2 \Phi_1^\dagger\Phi_2$ in the N2HDM. In the 2HDMS mass basis, see \refeq{eq:inpmass}, there is no free input parameter as $m_{12}^2$ due to the \zet3 symmetry. However, the term $\sim\Phi_1^\dagger\Phi_2$ is generated by $v_S$, the CP-odd Higgs masses and mixing angle, i.e.\  $\mu_{12}$ can be converted to a combination of $m_{a_1}$, $m_{a_2}$ and $\alpha_4$. 

For our analysis we interpret the experimental excess at $\sim 96 \gev$ as the lightest scalar Higgs boson $h_1$, and we identify the second lightest scalar Higgs $h_2$ as the SM-like Higgs at $\sim 125 \gev$.

Furthermore, the elements of the rotation matrix, i.e. $|R_{ij}|^2$, represent each field admixture of the corresponding physical state. The matrix elements thus determine the Higgs-bosons couplings to the SM particles. Here we define the reduced coupling as the ratio between the 2HDMS/N2HDM Higgs coupling and the corresponding SM-Higgs coupling:
\begin{equation}
	c_{h_i ff} = \frac{g_{h_i ff}}{g_{H_\text{SM}ff}}~.
\end{equation}
The reduced Higgs to fermion couplings for all four Yukawa types are summarized in \refta{tab:fermioncoup}.

\begin{table}[htb]
	\centering
\renewcommand{\arraystretch}{1.6}
	\begin{tabular}{l|c|c|c|c}
		\hline
		&type I& type II& lepton-specific& flipped\\
		\hline
		$c_{h_i tt}$& $\frac{R_{i2}}{\sin\beta}$& $\frac{R_{i2}}{\sin\beta}$& $\frac{R_{i2}}{\sin\beta}$& $\frac{R_{i2}}{\sin\beta}$\\
		$c_{h_i bb}$& $\frac{R_{i2}}{\sin\beta}$& $\frac{R_{i1}}{\cos\beta}$& $\frac{R_{i2}}{\sin\beta}$& $\frac{R_{i1}}{\cos\beta}$\\$c_{h_i \tau\tau}$& $\frac{R_{i2}}{\sin\beta}$& $\frac{R_{i1}}{\cos\beta}$& $\frac{R_{i1}}{\cos\beta}$& $\frac{R_{i2}}{\sin\beta}$\\
		\hline
	\end{tabular}
	\caption{Higgs to fermion reduced couplings for different type of Yukawa couplings}
	\label{tab:fermioncoup}
\renewcommand{\arraystretch}{1.2}
\end{table}

\noindent
One can also derive the reduced Higgs to gauge-bosons couplings,
\begin{equation}
	c_{h_iVV}=c_{h_iZZ}=c_{h_iWW}=\cos\beta R_{i1}+\sin\beta R_{i2}
	\label{eq:bosoncoup}
\end{equation}


\subsection{Type II SM-like Higgs boson}
\label{sec:SM-higgs}

Following \refeq{eq:rot}, one finds that the singlet component $h_1$ can be expressed by $|R_{13}|^2=\sin^2\alpha_2$. In our study, the lightest scalar Higgs $h_1$ should be a singlet-dominant Higgs, which is motivated by the experimental excesses, see the discussion below, i.e.\ $\sin^2\al_2$ should be close to 1.

Since the type~II N2HDM is favoured for interpreting the experimental excess~\cite{Biekotter:2019kde,Biekotter:2019mib,Biekotter:2019drs,Biekotter:2020ahz,Biekotter:2020cjs}, we will stick to the type~II Yukawa structure for our analysis. We choose $h_2$ to be the SM-like Higgs, and one can obtain the reduced couplings of $h_2$ to $t$-quark, $b$-quark and gauge bosons from \refta{tab:fermioncoup}, \refeq{eq:rot} and \refeq{eq:bosoncoup}, 
\begin{align}
c_{h_2tt}&=(c_{\alpha_1}c_{\alpha_3}-s_{\alpha_1}s_{\alpha_2}s_{\alpha_3})/s_\beta~,\\	c_{h_2bb}&=(-s_{\alpha_1}c_{\alpha_3}-c_{\alpha_1}s_{\alpha_2}s_{\alpha_3})/c_\beta~,\\	
	c_{h_2VV}&=c_{\alpha_3}s_{\beta-\alpha_1}-s_{\alpha_2}s_{\alpha_3}c_{\beta-\alpha_1}~.
\end{align}
In the limit of $\sin^2\alpha_2\rightarrow 1$, one can factor out  $|\sin\alpha_2|$, and the $h_2$ couplings are approximately givn by,
\begin{align}
c_{h_2tt}&\approx \frac{\cos(\alpha_1+\sgn(\alpha_2)\alpha_3)}{\sin\beta}|\sin\alpha_2|~,\label{h2ttcouplings}\\
c_{h_2bb}&\approx-\frac{\sin(\alpha_1+\sgn(\alpha_2)\alpha_3)}{\cos\beta}|\sin\alpha_2|~,\label{h2bbcouplings}\\
c_{h_2VV}&\approx \sin(\beta- (\alpha_1+\sgn(\alpha_2)\alpha_3))|\sin{\alpha_2}|~.\label{h2vvcouplings}
\end{align}
These three reduced couplings, \refeqs{h2ttcouplings} - (\ref{h2vvcouplings}), are required to be close to~1 for an SM-like $h_2$. The so-called alignment limit is thus reached for $\be - (\al_1 + \sgn(\al_2)\al_3) \to \pi/2$. 
All three couplings of the $h_2$ are close to 1 simultaneously in this limit.


\subsection{Differences between the 2HDMS and the N2HDM}
\label{sec:model}

As discussed above, the 2HDMS has an additional CP-odd Higgs-boson and  additional mixing angle $\alpha_4$ compared to the N2HDM, because of the imaginary part of the singlet field. Therefore, the $\alpha_4$ determines whether the lighter $a_1$ or the heavier $a_2$ plays the role of the singlet-like CP-odd Higgs. By taking our convention for the CP-odd mixing matrix in \refeq{eq:rota}, the lighter CP-odd Higgs $a_1$ would be singlet dominant when $\alpha_4\rightarrow\pi/2$. Conversely, when $\alpha_4\rightarrow 0$, the $a_1$ would be the doublet-like and the heavier $a_2$ become singlet-like. If $\alpha_4=\pi/4$, both $a_1$ and $a_2$ are the admixture of the singlet component and the doublet components. In case the $\alpha_4$ is exactly 0 or $\pi/2$, the singlet dominant CP-odd Higgs would completely decouple from the other SM particles, where the 2HDMS can be approximately in the "N2HDM limit". 

However, even in this limit the two models differ by their symmetries. The $\zet3$ symmetry of the 2HDMS yields two additional trilinear terms $\mu_{12}$ and $\mu_{S1}$ in the Higgs potential. By neglecting the effect of the imaginary part of the singlet field, the $\lambda'_1, \lambda'_2$ and $\lambda''_3$ in \refeq{eq:2hdmspot} can play the similar roles of $\lambda_7,\lambda_8$ and $\lambda_6$ in \refeq{eq:n2hdmpot}, respectively. On the other hand, the terms given by $\mu_{S1}$ and $\mu_{12}$ have no corresponding terms in the N2HDM. Consequently, these two terms can give additional contribution on the triple-Higgs couplings which can be expressed as,
\begin{align}
        \lambda_{h_ih_jh_k} &= \lambda_{h_ih_jh_k}^\text{N2HDM-like}
        +\frac{\mu_{S1}}{2v} R_{i3}R_{j3}R_{k3} \\
        &+\frac{\mu_{12}}{2v}[(R_{i2}R_{j3}+R_{j2}R_{i3})R_{k1} +(R_{i1}R_{j3}+R_{j1}R_{i3})R_{k2}+(R_{i1}R_{j2}+R_{j2}R_{i1})R_{k3}]~. \non
\label{eq:triHiggs}
\end{align}
Here the N2HDM-like part is the N2HDM triple Higgs couplings, but replacing the $\lambda_7,\lambda_8,\lambda_6$ by the $\lambda'_1, \lambda'_2,\lambda''_3$. Overall, the additional contributions can lead to the differences in $\la_{h_3h_ih_j}$ and thus in $\Ga(h_3\rightarrow h_i h_j)$.


\section{The constraints}
\label{sec:constraints}

In this section we describe in detail the theoretical and experimental
constraints applied in our analysis to the 2HDMS and N2HDM.


\subsection{Theoretical constraints}
\label{sec:theo}

The 2HDMS and N2HDM face constraints from tree-level perturbative unitarity, the condition that the potential should be bounded from below and the stability of the vacuum. In the following we show the conditions for the 2HDMS. All constraints for the N2HDM were already derived in~\cite{Muhlleitner:2016mzt} (see below). 

\begin{itemize}

\item \textbf{Tree-Level perturbative unitarity}

Tree-Level perturbative unitarity conditions ensures perturbativity of the model up to very high scales. This can be achieved by demanding the amplitudes of the scalar quartic interactions, which are given by the eigenvalues of the 2 $\rightarrow$ 2 scattering matrix, to be below a value of $8\pi$. The calculation was carried out with a \Code{Mathematica} package implemented in \Code{ScannerS}~\cite{scanners} and by following the procedure of~\cite{Ho_ej__2006}. The conditions are:

\begin{align}
|\lambda_{1,2}^{\prime}| &< 8\pi \\
|\frac{\lambda_3^{\prime\prime}}{2}| &< 8\pi \\
|\lambda_{1,2,3}| &< 8\pi \\
|\lambda_3\pm\lambda_4| &< 8\pi \\
|\frac{1}{2}(\lambda_1+\lambda_2\pm\sqrt{(\lambda_1-\lambda_2)^{2}+4\lambda_4^2})| &< 8\pi \\
\end{align}

For models with extended scalar-sectors the calculation cannot be carried out purely analytical. The remaining eigenvalues are given by the three real roots ($x_1, x_2, x_3$) of the cubic polynomial

\begin{gather}
\begin{split}
&64\,(6 \lambda_2^{\prime 2} \lambda_1 + 6 \lambda_1^{\prime 2} \lambda_2 - 
9 \lambda_3^{\prime\prime} \lambda_1 \lambda_2 - 8 \lambda_1^{\prime} \lambda_ 2^{\prime} \lambda_3 + 
4 \lambda_3^{\prime\prime} \lambda_3^2 - 4 \lambda_1^{\prime} \lambda_2^{\prime} \lambda_4 + 
4 \lambda_3^{\prime\prime} \lambda_3 \lambda_4 + \lambda_3^{\prime\prime} \lambda_4^2) \\ 
&+16\,(-2 \lambda_1^{\prime 2} - 2\lambda_2^{\prime 2} + 3 \lambda_3^{\prime\prime} \lambda_1 + 3 \lambda_3^{\prime\prime} \lambda_2 + 9 \lambda_1 \lambda_2 - 4 \lambda_3^2 - 4 \lambda_3 \lambda_4 - \lambda_4^2)\,x \\
&+ (-4 \lambda_3^{\prime\prime} - 12 \lambda_1 - 12 \lambda_2)\,x^2 +\,x^3 = 0
\end{split}\\
\Big| \frac{x_1}{4} \Big| < 8\pi,\qquad \Big| \frac{x_2}{4} \Big| < 8\pi,\qquad \Big| \frac{x_3}{4} \Big| < 8\pi 
\label{eq:uni_poly}
\end{gather}

The corresponding conditions for the N2HDM were already derived in~\cite{Muhlleitner:2016mzt} (see their Eqs.\ (3.43)-(3.48)).

\item \textbf{Boundedness from below}

The boundedness from below conditions ensures that the potential remains positive when the field values approach infinity.
The conditions can be found in~\cite{klimenko} and were adapted for the 2HDMS. The allowed region is given by

\begin{align}
\Omega_1 \cup \Omega_2
\end{align}

with

\begin{align}
\Omega_1 = \bigg\{&\lambda_1,\lambda_2,\lambda_3^{\prime\prime} > 0; \sqrt{\frac{\lambda_1\lambda_3^{\prime\prime}}{2}} + \lambda_1^{\prime} > 0; \sqrt{\frac{\lambda_2\lambda_3^{\prime\prime}}{2}} + \lambda_2^{\prime} > 0; \\
&\sqrt{\lambda_1\lambda_2}+\lambda_3+D > 0; \lambda'_1+\sqrt{\frac{\lambda_1}{\lambda_2}}\lambda_2^{\prime} \geq 0  \bigg\}
\end{align}

and

\begin{align}
\Omega_2 = \bigg\{&\lambda_1,\lambda_2,\lambda_3^{\prime\prime} > 0; \sqrt{\frac{\lambda_2\lambda_3^{\prime\prime}}{2}} \geq \lambda_2^{\prime} > -\sqrt{\frac{\lambda_2\lambda_3^{\prime\prime}}{2}}; -\sqrt{\frac{\lambda_1}{\lambda_2}}\lambda_2^{\prime} \geq \lambda_1^{\prime} > -\sqrt{\frac{\lambda_1\lambda_3^{\prime\prime}}{2}}; \\
&\frac{(D+\lambda_3)\lambda_3^{\prime\prime}}{2} > \lambda_1^{\prime}\lambda_2^{\prime} - 
\sqrt{(\lambda_1^{\prime 2}-\frac{\lambda_1\lambda_3^{\prime\prime}}{2})(\lambda_2^{\prime 2}-\frac{\lambda_2\lambda_3^{\prime\prime}}{2})} 
\bigg\},
\end{align}

where 

\begin{align}
D = 	
\left\{
\begin{matrix}
\lambda_4 \quad \text{for}\, \lambda_4 < 0 \\
0 \quad \text{for}\, \lambda_4 \geq 0\\
\end{matrix} \quad .
\right. 
\end{align}

The corresponding conditions for the N2HDM were already derived in~\cite{Muhlleitner:2016mzt} (see their Eqs.\ (3.51) and (3.52)).


\item \textbf{Vacuum stability}

In the SM the Electroweak (EW) vacuum is required to be stable at the EW scale. This vacuum state is characterised by a non-zero vev of the Higgs field. In BSM theories vacuum stability at the EW scale places additional constraints on their extended parameter space. An obvious condition is to require the EW vacuum to be the global minimum (\textit{true vacuum}) of the scalar potential. In this case the EW is absolutely stable. If the EW vacuum is a local minimum (\textit{false vacuum}) the corresponding parameter region can still be allowed if it is sufficiently metastable. This is the case if the predicted life-time of the false vacuum is longer than the current age of the universe. Any configuration with a life-time shorter than the age of the universe is considered unstable. 

For our study we used \Code{EVADE}~\cite{Ferreira:2019,Hollik:2019,evade:online} which finds the tree-level minima employing \Code{HOM4PS2}~\cite{lee2008hom4ps}. In the case of the EW  vacuum being a false vacuum, it calculates the bounce action for a given parameter point with a straight path approximation, which is sufficiently accurate for the purpose, see \cite{Hollik:2019}.
Points with a bounce action $B > 440$ are considered to be long-lived. We compared the results of the straight path approximation of \Code{EVADE} with the more sophisticated approach via path deformation of the code \Code{FindBounce}~\cite{findbounce2020}. We found the enhancement by the computationally more intensive \Code{FindBounce} to be negligible. Additionally we compared the results of \Code{EVADE} with the code \Code{Vevacious++}~\cite{vevaciouspp,vevacious}. The latter incorporates the tree-level as well as the Coleman-Weinberg one-loop potential. While the one-loop effective potential approach of \Code{Vevacious++} 
suffers from numerical instabilities, the tree-level results placed a stricter constraint on the long-lived regions, but where in good agreement with \Code{EVADE}. We chose the active developed code \Code{EVADE} over \Code{Vevacious++} which is based on the relatively outdated \Code{Vevacious}~\cite{vevacious}.

\end{itemize}


\subsection{Experimental Constraints}
\label{sec:expconstrain}

The set of experimental constraints from searches at colliders are the same for both the 2HDMS and the N2HDM.

\begin{itemize}

\item \textbf{Higgs-boson rate measurements}

We use the public code \texttt{HiggsSignals-2.6.1}~\cite{Bechtle:2013xfa,Bechtle:2014ewa,Bechtle:2008jh,Bechtle:2011sb,Bechtle:2013wla,Bechtle:2015pma}, to verify that all generated points agree with currently available measurements of the SM Higgs-boson. \Code{HiggsSignals} calculates the $\chi^2_{\text{HS}}$ from the comparison of the model prediction with the Higgs-boson signal rates and masses at Tevatron and LHC. The complete list of implemented constraints can be found in \cite{Bechtle:2013xfa}. In our analysis we use the reduced $\chi^2$ to judge the validity of our generated points, which is defined as,
\begin{align}
\chi^2_{\text{red}} = \frac{\chi^2_{\text{HS}}}{n_{\text{obs}}}~.
\label{eq:chired}
\end{align}
Here $\chi^2$ is evaluated by \Code{HiggsSignals} and $n_{\text{obs}}=111$ is the considered number of experimental measurements.


\item \textbf{BSM Higgs-boson searches}

The public code \texttt{HiggsBounds-5.9.1}~\cite{Bechtle:2008jh,Bechtle:2011sb,Bechtle:2013wla,Bechtle:2015pma} provides $95\%$ confidence level exclusion limits of all relevant direct searches for charged Higgs bosons and additional neutral Higgs-bosons. Searches for charged Higgs-bosons put constraints on the allowed $M_{H^\pm}-\text{tan}\beta$ regions in the 2HDM~\cite{Arbey:2017gmh}.  Since the charged sector is identical in the 2HDMS the constraints on the parameter space can be directly taken over. Important searches are the direct searches for charged Higgs production $pp \rightarrow H^\pm t b$ with the decay modes $H^\pm \rightarrow \tau \nu$ and $H^\pm \rightarrow t b$~\cite{Atlas:13TeVHtb}. The constrained regions mostly lie in the low tan$\beta \lsim 2$ region, due to the enhanced coupling to top quarks. Searches at LEP for charged Higgs-bosons are mostly irrelevant as we focus on tan$\beta = \{1,20\}$ and light charged Higgs-boson masses are excluded from flavour physics observables (see below).

Direct searches for additional neutral Higgs-bosons are relevant when the heavy scaler Higgs-boson $h_3$ or the heavy pseudoscalar Higgs-bosons $a_1, a_2$ are not too heavy.


\item \textbf{Flavour physics observables}

The presence of additional Higgs Bosons in 2HDM-type models leads to constraints from flavour physics.
The charged Higgs sector of the 2HDMS or N2HDM is unaltered with respect to the general 2HDM. For the tan$\beta$ = $\{1,20\}$ we are interested in, the most important bounds according to~\cite{Arbey:2017gmh} come from BR$(B_s\rightarrow X_s \gamma)$, constraints on $\Delta M_{B_s}$ from neutral B-meson mixing and BR$(B_s\rightarrow\mu^+\mu^-)$. The dominant contributions to these bounds come from charged Higgs $H^\pm$~\cite{bxs1,bxs2,bxs3} and top quarks~\cite{flavor1,flavor2}. As they are independent from the neutral scalar sector to a good approximation we can take over the bounds directly from the 2HDM. The constraints from $\Delta M_{B_s}$ and BR$(B_s\rightarrow\mu^+\mu^-)$ are dominant for tan$\beta \simeq 1$ while the constraint from BR$(B_s\rightarrow X_s \gamma)$ is present for the whole range of tan$\beta$ we study. Taking all this into account for our study in the type II 2HDMS and N2HDM, these constraints give a lower limit of the charged Higgs mass of $\MHp \gsim 650 \gev$~\cite{Arbey:2017gmh}.


\item \textbf{Electroweak precision observables}

 If BSM physics enter mainly through gauge boson self-energies, as it is the case for the extended Higgs sector of the 2HDMS and the N2HDM, constraints from electroweak precision observables can be expressed in terms of the parameters 
 $S$, $T$ and $U$~\cite{Grimus:2008}, revealing significant changes from BSM to these parameters. SPheno has implemented the one-loop corrections of these parameters for any model with additional doublets and singlets. If the gauge group is the SM $SU(2)\, \times \,U(1)$ and BSM particles have suppressed couplings to light SM fermions, the corrections are independent of the Yukawa type of the model. The differences of the masses of the scalars have a large impact on these corrections. They are small if the heavy Higgs mass $h_3$ or the heavy doublet-like pseudo-scalar mass are close to the charged Higgs-boson mass~\cite{Grimus:2008}. In the 2HDMS and the N2HDM the parameter $T$ is the most sensitive and has a strong correlation to $U$. Contributions to $U$ can therefore be dropped~\cite{Biekotter:2019kde}. For our scan we require the prediction of the $S$ and $T$ parameters to be within the 95\%~CL region, corresponding to $\chi^2 = 5.99$ for two degrees of freedom.

\end{itemize}
\section{The parameter space embedding the experimental excess}
\subsection{The 96 GeV "excesses"}
\label{sec:excesses}
The experimental excesses at both LEP and CMS could be translated to the following signal strengths as quoted in \cite{Cao:2016uwt, Schael:2006cr, Sirunyan:2018aui, Gascon:2017hd}:
\begin{align}
	\mu^\text{exp}_\text{LEP}&=\frac{\sigma^{{\text{exp}}}(e^+ e^-\rightarrow Z \phi\rightarrow Z b \bar{b})}{\sigma^{{\text{SM}}}(e^+ e^-\rightarrow Z H^0_\text{SM}\rightarrow Z b \bar{b})}=0.117\pm0.05\\
	\mu^\text{exp}_\text{CMS}&=\frac{\sigma^{{\text{exp}}}(p p\rightarrow \phi\rightarrow \gamma\gamma)}{\sigma^{{\text{SM}}}(p p\rightarrow H^0_\text{SM}\rightarrow \gamma\gamma)}=0.6\pm0.2
\end{align}
where the $H^0_\text{SM}$ is the SM Higgs-boson with the rescaled mass at the same range as the unknown scalar particle $\phi$.

Since one of the most important targets of our analysis is the interpretation of the experimental excess in the 2HDMS{/N2HDM}, we interpreted the scalar $\phi$ as the lightest CP-even Higgs-boson $h_1$ of the 2HDMS{/N2HDM}, and we evaluate such signal strengths for all the $h_1$. These signal strengths can be calculated by the following expressions in the narrow width approximation \cite{Biekotter:2019kde} {({introduced} here for the 2HDMS)}:
\begin{align}
	\mu^\text{the}_\text{LEP}&=\frac{\sigma_\text{2HDMS}(e^+ e^-\rightarrow Z h_1)}{\sigma_\text{SM}(e^+ e^-\rightarrow Z H^0_\text{SM})}\times\frac{\text{BR}_\text{2HDMS}(h_1\rightarrow b\bar{b})}{\text{BR}_\text{SM}(H^0_\text{SM}\rightarrow b\bar{b})}=|c_{h_1 VV}|^2\frac{\text{BR}_\text{2HDMS}(h_1\rightarrow b\bar{b})}{\text{BR}_\text{SM}(H^0_\text{SM}\rightarrow b\bar{b})}
	\label{eq:mulep}
	\end{align}
	\begin{align}
	\mu^\text{the}_\text{CMS}&=\frac{\sigma_\text{2HDMS}(g g\rightarrow h_1)}{\sigma_\text{SM}(g g\rightarrow H^0_\text{SM})}\times \frac{\text{BR}_\text{2HDMS}(h_1\rightarrow \gamma\gamma)}{\text{BR}_\text{SM}(H^0_\text{SM}\rightarrow \gamma\gamma)}=|c_{h_1 tt}|^2\frac{\text{BR}_\text{2HDMS}(h_1\rightarrow \gamma\gamma)}{\text{BR}_\text{SM}(H^0_\text{SM}\rightarrow \gamma\gamma)}
	\label{eq:mucms}
\end{align}
The effective couplings of $c_{h_1 VV}$ and $c_{h_1 tt}$ can be easily obtained from \refeq{eq:bosoncoup} and \refta{tab:fermioncoup}, while the {corresponding branching ratios have been obtained with} \texttt{SPheno-4.0.4}~\cite{Porod:2003um,Porod:2011nf}.

{The overall $\chi^2$ corresponding to the excesses is calculated as,}
\begin{equation}
	\chi_\text{CMS-LEP}^2=\left(\frac{\mu_\mathrm{LEP}^\text{the}-0.117}{0.057}\right)^2+\left(\frac{\mu_\mathrm{CMS}^\text{the}-0.6}{0.2}\right)^2
	\label{eq:chi2excess}
\end{equation}
{The points of the 2HDMS/N2HDM with the lowest $\chi^2$ are the respective "best-fit" points in the two models.}

\medskip
In order to understand the effect of mixing angles on the signal strengths of the excesses, one can focus on the couplings of $h_1$ derived from \refeq{eq:rot} and \refta{tab:fermioncoup}, which are given by:
\begin{equation}
	c_{h_1 tt}=\frac{\sin\alpha_1\cos\alpha_2}{\sin\beta},\qquad c_{h_1 bb}=\frac{\cos\alpha_1\cos\alpha_2}{\cos\beta},\qquad c_{h_1VV}=\cos\alpha_2\cos(\beta-\alpha_1)~.
	\label{eq:h1coup}
\end{equation}
If $h_1$ is the pure gauge singlet (i.e. $\cos\alpha_2=0$), all the three couplings in \refeq{eq:h1coup}, which are proportional to the $\cos\alpha_2$, would be zero. However, $h_1$ would then be completely invisible and could not produce any experimental excesses in this case. Therefore, in order to cover the ranges of the experimental excesses efficiently, we enforced the singlet component of $h_1$ to be smaller than 95\% (i.e.\ $\cos^2\alpha_2>5\%$), which essentially yields non-vanishing couplings of $h_1$ to SM particles. As a result, the interval of $\alpha_2$ is constrained by this requirement. We have checked explicitly that this constraint does not exclude any valid parameter point in our analysis.

For the signal strength of CMS, the coupling $c_{h_1tt}$ and the $\text{BR}(h_1\rightarrow \gamma\gamma)$ play the dominant roles. Since the decay width of the $h_1$ is dominated by the decay to $b\bar b$, a smaller $c_{h_1 bb}$ would suppress the decay width of $h_1\rightarrow b\bar{b}$ and lead to the enhancement of $\text{BR}(h_1\rightarrow \gamma\gamma)$. Consequently, $\text{BR}(h_1\rightarrow \gamma\gamma)$ can be anti-proportional to the coupling ${|c_{h_1bb}|^2}$. Since $\mu_\text{CMS}$ is also proportional to the $|c_{h_1tt}|^2$, one obtains the approximate relation for $\mu_\text{CMS}$ which is given by:
\begin{equation}
\mu^\text{the}_\text{CMS}\propto \frac{|c_{h_1tt}|^2}{|c_{h_1bb}|^2}=\left(\frac{\tan\alpha_1}{\tan\beta}\right)^2~.
\label{eq:mucmsappr}
\end{equation}
As we see in the \refeq{eq:mucmsappr}, $\mu^\text{the}_\text{CMS}$ can be directly enhanced by the increment of $\alpha_1$. In order to have a not too suppressed signal strength for the CMS excess, $\tan\alpha_1>\tan\beta$ is required. However, the combination $\frac{\tan\alpha_1}{\tan\beta}$ can be arbitrarily large during the scan. Thus we scan the inverse of this combination in the range from 0 to 1, see the next subsection.


\subsection{Parameter scan}
\label{sec:scan}

Following the N2HDM interpretation of the excesses~\cite{Biekotter:2019kde}, we focus on the type-II Yukawa structure also for the 2HDMS. However, we will investigate a larger $\tb$ region as it was done in~\citere{Biekotter:2019kde}.

In order to investigate the parameter space of the 2HDMS/N2HDM that gives rise to a description of the $96 \gev$ excesses, we performed an extensive scan of the parameter spaces by using the spectrum generator \texttt{SPheno-4.0.4}~\cite{Porod:2003um, Porod:2011nf}, where the model implementations are generated by the public code \texttt{SARAH-4.14.3}~\cite{Staub:2013tta}. During the scan, we fix the mass $m_{h_2}=125.09 \gev$ and enforce the mixing angles to be close to the alignment limit as explained in detail in \refse{sec:SM-higgs}. As discussed above, by employing \texttt{HiggsSignal-2.6.1}, we can ensure that the $h_2$ is in agreement with the LHC measurements. Concerning the exclusion bounds from flavor physics as we mentioned in \refse{sec:expconstrain}, we simply apply the conservative limits given by $\tan\beta>1$ and $m_{H^\pm}>800 \gev$, which is above the experimental limit of $650 \gev$~\cite{Arbey:2017gmh}. 

In our study for simplicity we assumed that the lightest CP-odd Higgs $a_1$ is singlet dominated, which corresponds to the condition of $|R^A_{1,3}|^2 = \sin^2\alpha_4 > 1/2$, see \refeq{eq:rota}. In this case, the scan range of the $a_1$ mass is chosen from $200 \gev$ to $500 \gev$. Nevertheless, one could also explore the lower $m_{a_1}$ parameter space and open the decay channel of $h_2\rightarrow a_1 a_1$, but we leave this scenario for future studies.

Furthermore, we aim to study the parameter space in the region of $\tan\beta$ from 1 to 20 (i.\ e.\ going beyond the region explored in \citere{Biekotter:2019kde}). However, the lower bound of the heavy Higgs bosons masses $m_{h_3}$, $m_{a_2}$ and $m_{H^\pm}$ is raised as $\tan\beta$ goes to higher values because of the constraints from heavy Higgs-boson searches $H/A\rightarrow\tau^+\tau^-$ at the LHC\cite{Bagnaschi:2018ofa}. Therefore, we scan separately two $\tan\beta$ intervals with different ranges for the heavy Higgs-boson masses, see \refta{tab:high_tbscan}. Concerning the constraint of the unitarity and the $S,T,U$ parameters, the mass difference between the heavy Higgs states should be small and their scan intervals are therefore chosen to be identical. Overall, the scan intervals for all the particles are given by:
\begin{align}
	&m_{h_1}\in \{95,\;98\}\;\text{GeV}, &m_{h_2}&=125.09\;\text{GeV},&m_{a_1}&\in \{200,\;500\}\;\text{GeV},\notag\\
	&|\sin(\beta-(\alpha_1+\sgn(\alpha_2)\alpha_3))|\in \{0.98, 1\}&\alpha_4&\in \{\frac{\pi}{4},\;\frac{\pi}{2}\},&v_S&\in \{100,\;2000\}\,\text{GeV},\notag\\
	&\frac{\tan\beta}{\tan\alpha_1}\in \{0, 1\}, &\alpha_2&\in \pm\{0.95,\;1.3\}~. && 
	\label{eq:scanintervals}
\end{align}
\begin{table}[h]
	\centering
	\begin{tabular}{l|cc}
		\hline
		$\tan\beta$&  \;$m_{h_3} \sim m_{a_2} \sim m_{H^\pm}$\\
		\hline
		$1-10$&  $\{800,\;1200\}$ GeV\\
		$10-20$&  $\{1000,\;1700\}$ GeV\\
		\hline
	\end{tabular}
	\caption{Heavy Higgs-boson mass scan intervals for different $\tan\beta$ regions.}
	\label{tab:high_tbscan}
\end{table}
We have checked explicitly that the constraints on $\sin(\be - (\al_1 + \sgn(\al_2)\al3)$) and $\al_2$ do not exclude any valid point of our parameter space.

\subsection{Preferred 2HDMS parameter spaces}
\label{sec:pref2hdms}

\begin{figure}[h]
	\centering
 \includegraphics[width=.618\textwidth]{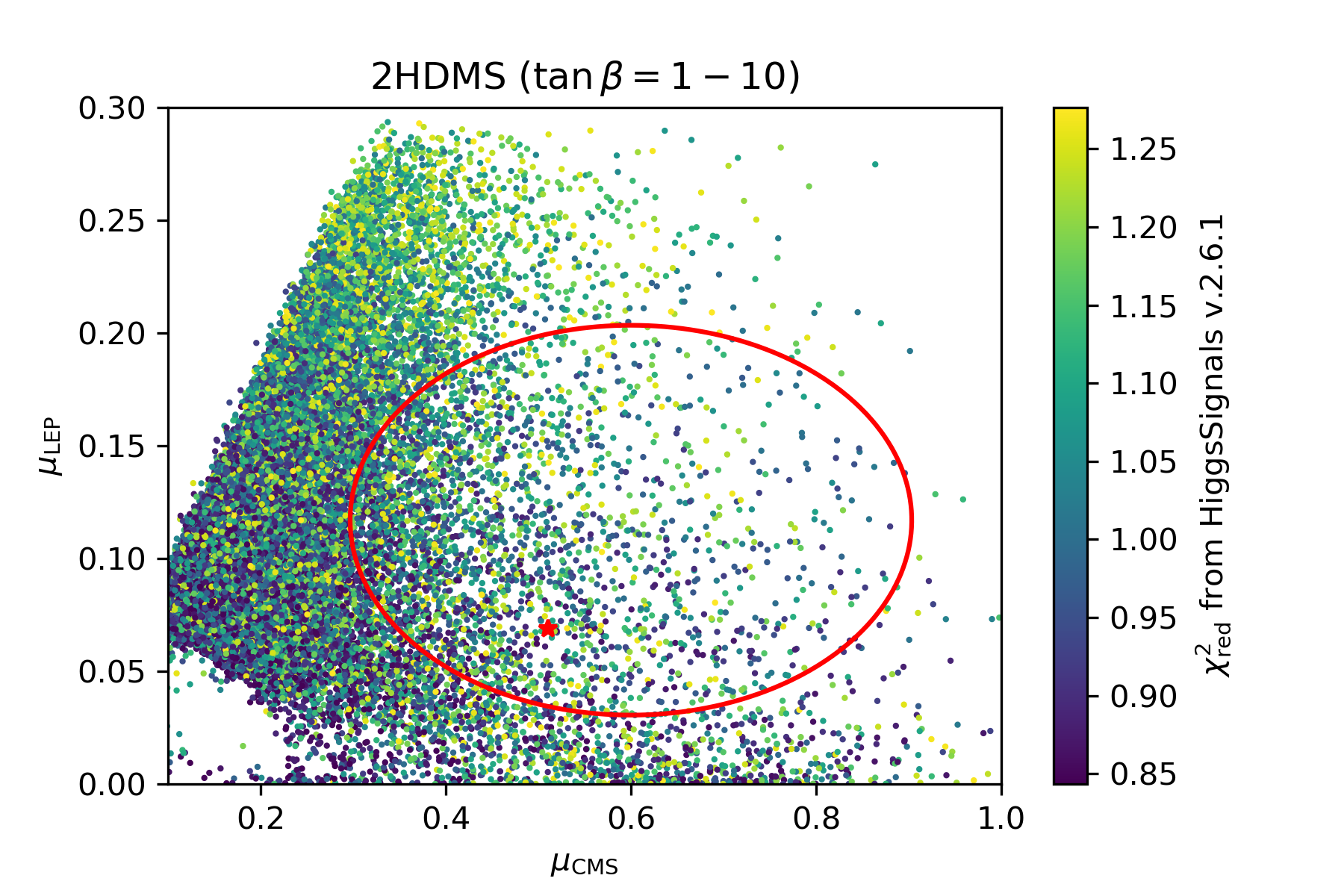}
 \caption{The signal strengths of both excesses $\mu_\text{CMS}$ and $\mu_\text{LEP}$ for the 2HDMS scan points with $\tan\beta\in\{1,\;10\}$. The red ellipse shows the $1\,\sig$ region of the excesses with the red star as the best-fit point. The color code indicates the $\chi^2_\text{red}$ and the lowest $\chi^2_\text{red}$ in the $1\,\sig$ ellipse is about 0.821.}
 \label{fig:excess}
\end{figure}

The results of the 2HDMS scan in the low-$\tb$ region are shown in \reffi{fig:excess} in the $\mu_{\text{CMS}}$-$\mu_{\text{LEP}}$ plane, where the color code indicates $\chi_{\rm red}^2$, see \refeq{eq:chired}. The red ellipse corresponds to the $1\,\sig$ ellipse, with the best-fit point (see below) marked by a red cross.

As can be seen in \reffi{fig:excess} the $1\,\sig$ ellipse of $\mu_\text{CMS}$ and $\mu_\text{LEP}$ can be fully covered by the 2HDMS parameter space, while all the points in the figure have  $\chi^2_\text{red}<1.3$. The lowest $\chi^2_\text{red}$ in the $1\,\sig$ ellipse is about 0.821.  Therefore, the $96 \gev$ excesses can be easily accommodated while the $h_2$ at $\sim 125 \gev$ is in good agreement with the experimental measurements. Combining the $\chi_\text{CMS-LEP}^2$ in \refeq{eq:chi2excess} and the $\chi^2_\text{HS}$ for the $125 \gev$ Higgs, we found the best-fit point which is given in \refta{tab:bestfit}.

\begin{table}[h]
	\centering\small
\begin{tabular}{cccccc}
	\hline
$m_{h_1}$&	$m_{h_2}$&	$m_{h_3}$&	$m_{a_1}$&	$m_{a_2}$&	$m_{H^\pm}$\\
96.438 GeV& 125.09 GeV& 784.08 GeV& 413.46 GeV& 660.07 GeV& 808.93 GeV\\
\hline
$\tan\beta$&	$\alpha_1$&	$\alpha_2$&	$\alpha_3$&	$\alpha_4$&	$v_s$\\
1.3393 & 1.3196 & -1.1687 & -1.2575 & 1.4719 & 653.84  GeV\\
\hline
\hline
\multicolumn{6}{c}{
Branching ratios
}\\
\hline
$h_1\rightarrow b\bar{b}$&   $h_1\rightarrow g g$&   $h_1\rightarrow\tau^+\tau^-$&    $h_1\rightarrow \gamma\gamma$&  $h_1\rightarrow {W^+}^{(*)} {W^-}^{(*)}$&   $h_1\rightarrow Z^{(*)}Z^{(*)}$\\
42.2\% & 35.3\% & 4.61\% & 0.317\% & 0.739\% & $<$ 0.1\%  \\
\hline
$h_2\rightarrow b\bar{b}$&   $h_2\rightarrow g g$&   $h_2\rightarrow\tau^+\tau^-$&    $h_2\rightarrow \gamma\gamma$&  $h_2\rightarrow {W^+}^{(*)} {W^-}^{(*)}$&   $h_2\rightarrow Z^{(*)}Z^{(*)}$\\
53.9\% & 10.5\% & 6.17\% & 0.249\% & 23.4\% & 2.54\%  \\
\hline
$h_3\rightarrow b \bar{b}$&   $h_3\rightarrow t\bar{t}$&   $h_3\rightarrow h_2 h_2$&    $h_3\rightarrow h_1 h_2$&  $h_3\rightarrow h_1 h_1$&   $h_3\rightarrow W^+ W^-$\\
$<$ 0.1\% & 65.3\% & 5.26\% & 7.76\% & 0.158\% & 8.26\%  \\
\hline
$a_1\rightarrow t \bar{t}$& $a_1\rightarrow \tau^+ \tau^-$& $a_2\rightarrow t\bar{t}$& $a_2\rightarrow \tau^+ \tau^-$& $H^\pm\rightarrow t b$& $H^\pm\rightarrow W^\pm h_2$\\
95\% & $<$ 0.1\% & 88.2\% & $<$ 0.1\% & 73.7\% & 1.12\%  \\
\hline
\end{tabular}
\caption{Parameters and relevant branching ratios of the best-fit point in the 2HDMS in the $\tan\beta\in\{1,\;10\}$ region. 
}
\label{tab:bestfit}
\end{table}

\begin{table}[htb!]
	\centering\small
\begin{tabular}{cccccc}
	\hline
	\multicolumn{6}{c}{Best fit point in $\tan\beta\in\{10,\;20\}$}\\
	\hline
$m_{h_1}$&	$m_{h_2}$&	$m_{h_3}$&	$m_{a_1}$&	$m_{a_2}$&	$m_{H^\pm}$\\
96.013 GeV& 125.09 GeV& 1437.8 GeV& 323.4 GeV& 1438.5 GeV& 1499.6 GeV\\
\hline
$\tan\beta$&	$\alpha_1$&	$\alpha_2$&	$\alpha_3$&	$\alpha_4$&	$v_s$\\
13.783 & 1.5441 & 1.2162 & 1.5338 & 1.5679 & 1212.6  GeV\\
\hline
\hline
\multicolumn{6}{c}{
Branching ratios
}\\
\hline
$h_1\rightarrow b\bar{b}$&   $h_1\rightarrow g g$&   $h_1\rightarrow\tau^+\tau^-$&    $h_1\rightarrow \gamma\gamma$&  $h_1\rightarrow {W^+}^{(*)} {W^-}^{(*)}$&   $h_1\rightarrow Z^{(*)}Z^{(*)}$\\
45.1\% & 32.5\% & 4.93\% & 0.612\% & 1.04\% & $<$ 0.1\%  \\
\hline
$h_2\rightarrow b\bar{b}$&   $h_2\rightarrow g g$&   $h_2\rightarrow\tau^+\tau^-$&    $h_2\rightarrow \gamma\gamma$&  $h_2\rightarrow {W^+}^{(*)} {W^-}^{(*)}$&   $h_2\rightarrow Z^{(*)}Z^{(*)}$\\
53.7\% & 10.0\% & 6.14\% & 0.269\% & 24.2\% & 2.62\%  \\
\hline
$h_3\rightarrow b \bar{b}$&   $h_3\rightarrow t\bar{t}$&   $h_3\rightarrow h_2 h_2$&    $h_3\rightarrow h_1 h_2$&  $h_3\rightarrow h_1 h_1$&   $h_3\rightarrow W^+ W^-$\\
69.7\% & 4.82\% & 3.74\% & 5.73\% & 0.585\% & 2.60\%  \\
\hline
$a_1\rightarrow b \bar{b}$& $a_1\rightarrow \tau^+ \tau^-$& $a_2\rightarrow b\bar{b}$& $a_2\rightarrow \tau^+ \tau^-$& $H^\pm\rightarrow t b$& $H^\pm\rightarrow W^\pm h_2$\\
88.0\% & 11.7\% & 74.2\% & 12\% & 91.4\% & 0.353\%  \\
\hline
\end{tabular}
\caption{Parameters and relevant branching ratios of the best-fit point in the high $\tb$ region.
}
\label{tab:bestfit2}
\end{table}

\begin{figure}[!]
\centering
\begin{minipage}{.48\textwidth}
	\centering
		\includegraphics[width=\textwidth]{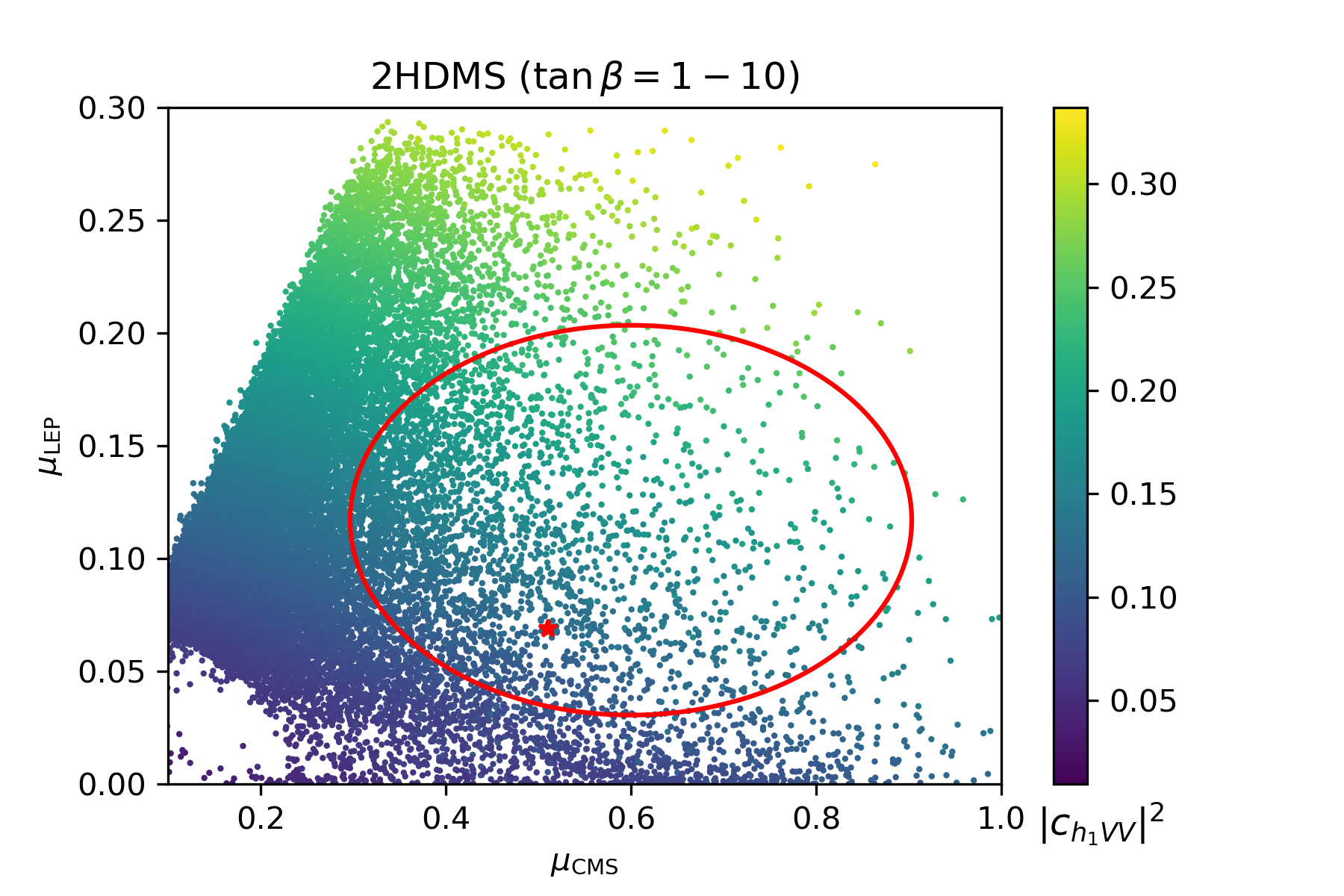}
	\caption{The same plane as in \protect\reffi{fig:excess}, with the color code indicating the square of the effective coupling of $h_1$ to gauge bosons. The lowest (highest) value of $|c_{h_1VV}|^2$ in the $1\,\sig$ ellipse is 0.088 (0.26).}
\label{fig:excess1}
\end{minipage}\quad
\begin{minipage}{.48\textwidth}
	\centering
\includegraphics[width=\textwidth]{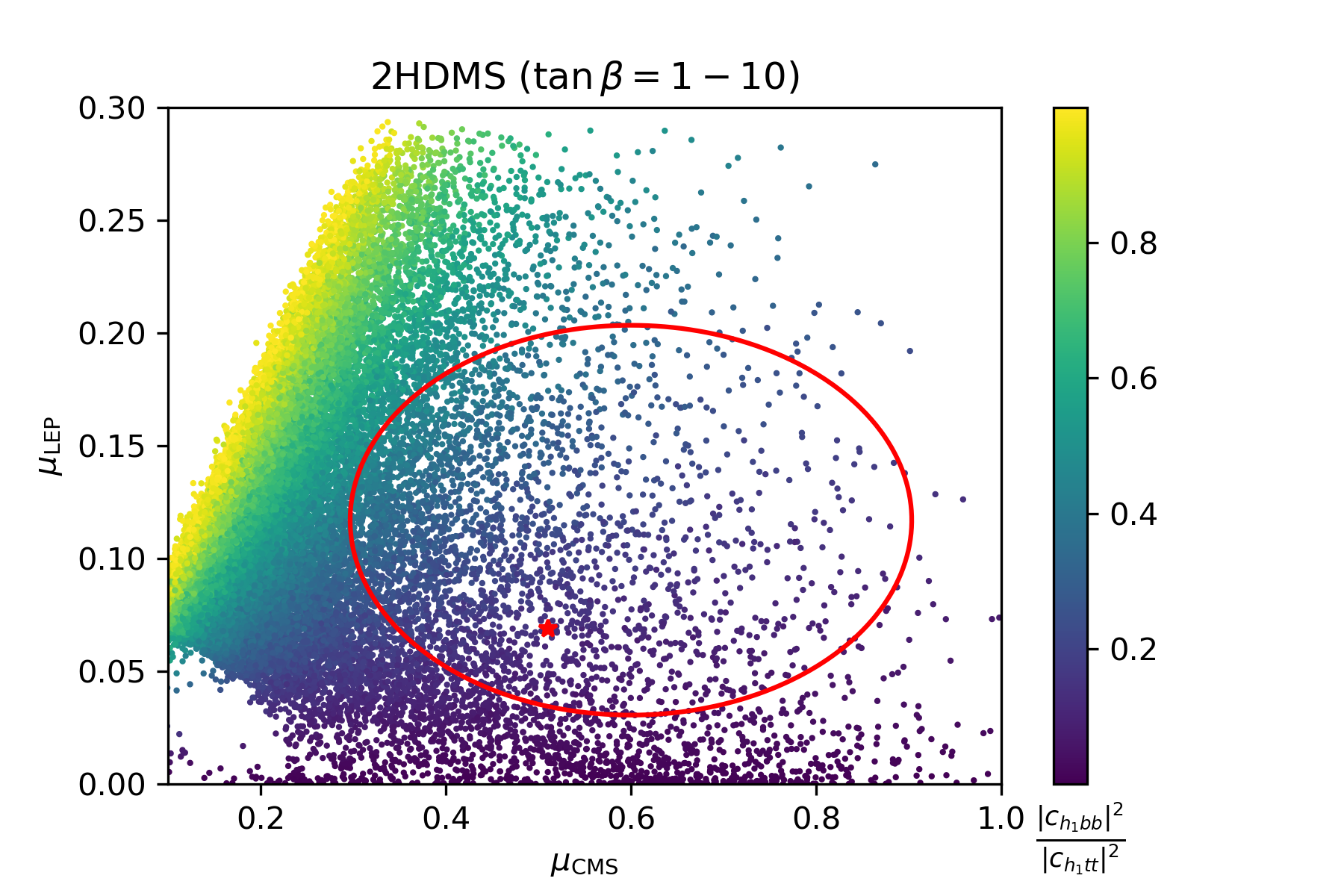}
	\caption{The same plane as in \protect\reffi{fig:excess}, with the color code indicating the ratio $|c_{h_1bb}|^2/|c_{h_1tt}|^2$. The lowest (hightest) value in the $1\,\sig$ ellipse is 0.039 (0.53).}
	\label{fig:excess2}
\end{minipage}
\end{figure}

In \reffis{fig:excess1}, \ref{fig:excess2} we show the results for $|c_{h_1VV}|^2$ and $|c_{h_1bb}/c_{h_1tt}|^2$, respectively, in the plane of $\mu_\text{CMS}$ and $\mu_\text{LEP}$. In \reffi{fig:excess1}, the points with higher signal strength $\mu_\text{LEP}$ always have the higher coupling $c_{h_1VV}$, as $\mu_\text{LEP}$ is directly proportional to $|c_{h_1VV}|^2$, see \refeq{eq:mulep}. On the other hand, one can observe from \reffi{fig:excess2} that the points with lower values of $|c_{h_1bb}|^2/|c_{h_1tt}|^2$ yield a higher signal strength $\mu_\text{CMS}$, consistent with the discussion in \refse{sec:excesses}, i.e.\ $\mu_\text{CMS}$ is anti-proportional to $|c_{h_1bb}|^2/|c_{h_1tt}|^2$. However, a lower $c_{h_1bb}$ coupling would slightly suppress $\br(h_1\rightarrow b\bar{b})$ that lead to the lower $\mu_\text{LEP}$, and therefore the distribution is slightly oblique in the plane of $\mu_\text{CMS}$ and $\mu_\text{LEP}$. The best fit point marked by the red cross has $|c_{h_1VV}|^2\sim0.13$ and $|c_{h_1bb}|^2/|c_{h_1tt}|^2\sim0.12$.

\begin{figure}[!]
	\centering
 \includegraphics[width=.618\textwidth]{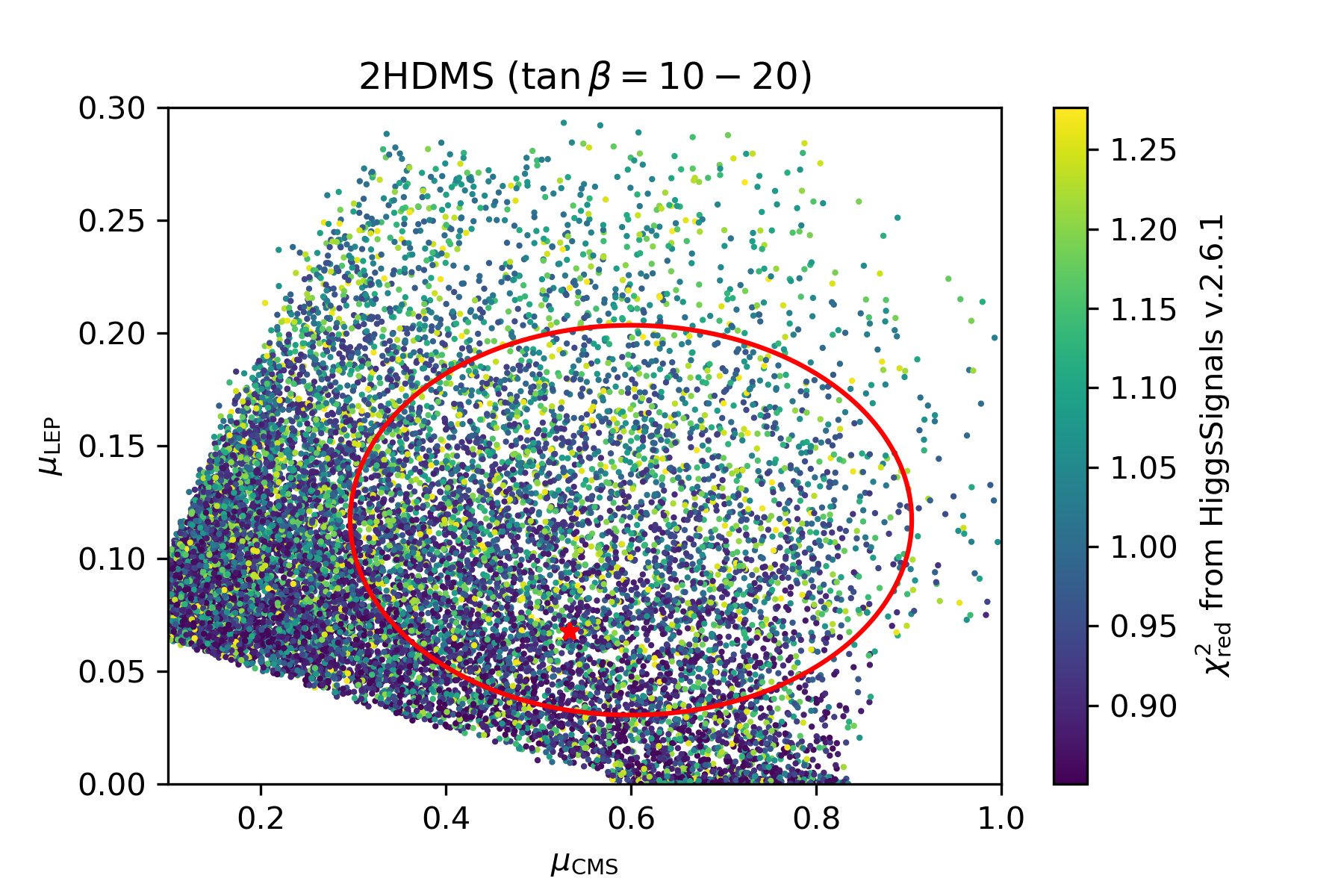}
 \caption{The same plane as in \protect\reffi{fig:excess} with $\tan\beta\in\{10,20\}$, and the color coding indicating the $\chi^2_{\text{red}}$.}
 \label{fig:excesstbhigh}
\end{figure}

The results of high $\tan\beta$ region scan, using the scan intervals given in \refta{tab:high_tbscan}, are shown in the \reffi{fig:excesstbhigh}, where the color coding indicates the $\chi^2_{\text{red}}$. It can be observed that also in the high $\tb$ region the $1\,\sig$ ellipse in the plane of $\mu_\text{CMS}-\mu_\text{LEP}$ is well covered by our parameter scan for $\tan\beta=10 - 20$. The distribution of the $\chi^2_{\text{red}}$ is found to be very similar to the low $\tb$ case. Also the other quantities, $|c_{h_1VV}|^2$ and $|c_{h_1bb}|^2/|c_{h_1tt}|^2$ behave as in the low $\tb$ case (and are thus not shown). In the \refta{tab:bestfit2} we summarize the details for best fit points in the region of $\tan\beta=10-20$. The high $\tan\beta$ best fit point  has $|c_{h_1VV}|^2 \sim 0.12$ and $|c_{h_1bb}/c_{h_1tt}|^2 \sim 0.14$, which is very close to the corresponding numbers of the low $\tan\beta$ best fit point. Overall we find that the points within the $1\,\sig$ range of the 96 GeV excesses have no preference for low or high $\tb$. Finally, also for the charged Higgs-boson mass we do not find a preferred region (within the intervals given in \refta{tab:high_tbscan}), neither in the low, nor in the high $\tb$ analysis. 


\subsection{Preferred N2HDM parameter spaces}
\label{sec:prefn2hdm}

\begin{figure}[htb]
	\centering
	\begin{subfigure}{.5\linewidth}
 \includegraphics[width=\textwidth]{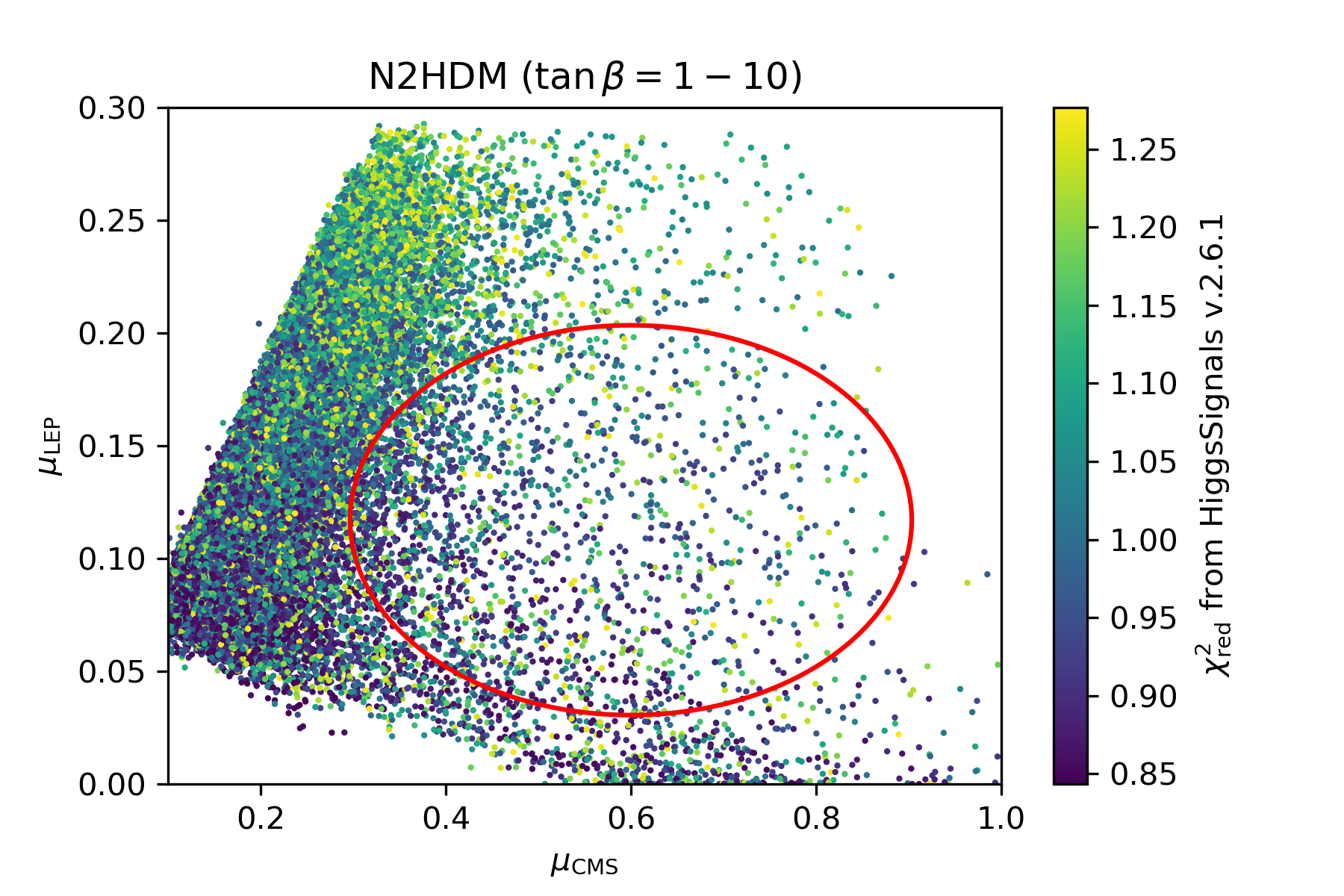}
	\end{subfigure}\begin{subfigure}{.5\linewidth}
 \includegraphics[width=\textwidth]{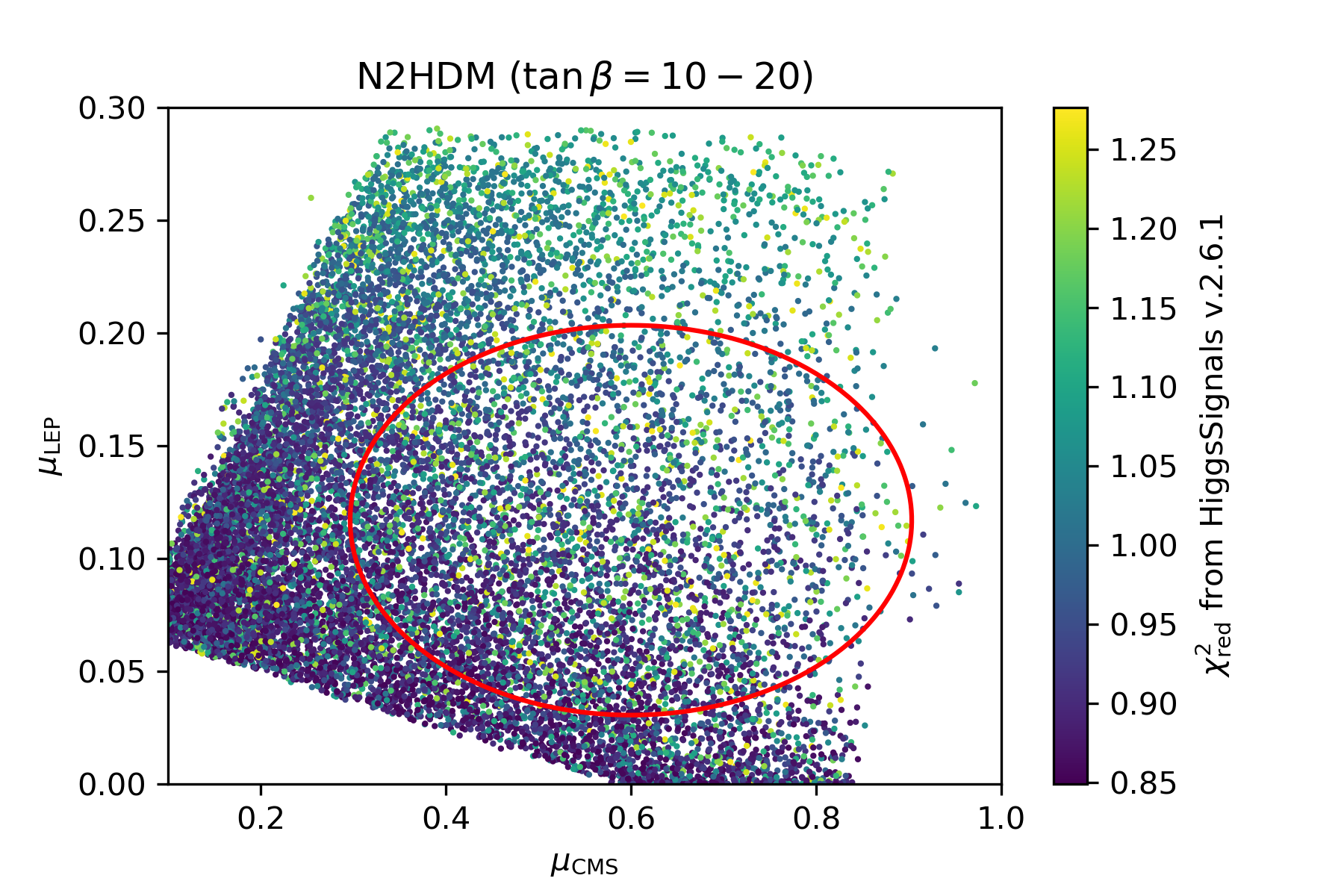}
	\end{subfigure}
	\caption{The N2HDM scan results with the same plane as in \protect\reffi{fig:excess}. The left (right) plot is for $\tan\beta=1-10$ ($10-20$).}
	\label{fig:n2hdmexcess}
\end{figure}

We now turn to the corresponding analysis in the N2HDM, where earlier results can be found in \citeres{Biekotter:2019kde,Biekotter:2019mib,Biekotter:2019drs,Biekotter:2020ahz,Biekotter:2020cjs}.
In \reffi{fig:n2hdmexcess} we show the results of the N2HDM in the $\mu_{\text{CMS}}$-$\mu_{\text{LEP}}$ plane for the low (left plot) and high $\tb$ range (right plot). One can observe that both the low $\tan\beta$ region and the high $\tan\beta$ region of the N2HDM parameter space can cover the $1\,\sig$ range of the 96 GeV "excess". This extends the analysis in \citere{Biekotter:2019kde}, where only relatively low $\tb$ values were found. These differences can be traced back to an improved scan strategy as well as improvements in the parameter point generation. The behavior of the other quantities analyzed in the previous subsection is very similar for the N2HDM.

Overall, we find that the 2HDMS and the N2HDM can fit equally well the $96 \gev$ excesses. The differences between the two models (different symmetries and different particle content) do not impact in a relevant way the description of the excesses.

\section{Prospects at the future colliders}
\label{sec:results}

The searches for a possible Higgs boson at $\sim 96 \gev$ will continue at ATLAS and CMS. However, it is not expected that such a particle could be seen in other decay modes than $\ga\ga$ and possibly $\tau^+\tau^-$. The $pp$ environment makes it difficult to perform precision measurements of such a light Higgs boson. Better suited for such a task would be a future $e^+e^-$ collider such as the planned ILC~\cite{Biekotter:2019kde,Biekotter:2020ahz}, where the light Higgs is produced in the Higgs-strahlung channel, $e^+e^- \to Z^* \to Z h_1$~\citeres{Drechsel:2018mgd,Wang:2018fcw,jenny}. The ILC can analyze the scenarios under investigations in two complementary ways. One can search for the new Higgs boson and analyze its properties directly. On the other hand, one can perform precision measurements of the Higgs-boson at $\sim 125 \gev$ and look for indirect effects of the extended Higgs-boson sector. In this section we will explore both possibilities (where we will emphasize where we go beyond \citeres{Biekotter:2019kde,Biekotter:2020ahz}. In particular, we analyze the 2HDMS and the N2HDM side-by-side to check for possible differences in the phenomenology.


\subsection{Direct search for the \boldmath{$96 \gev$ Higgs boson}}
\label{sec:ilc-h96}

\begin{figure}[htb]
    \centering
    \includegraphics[width=.7\textwidth]{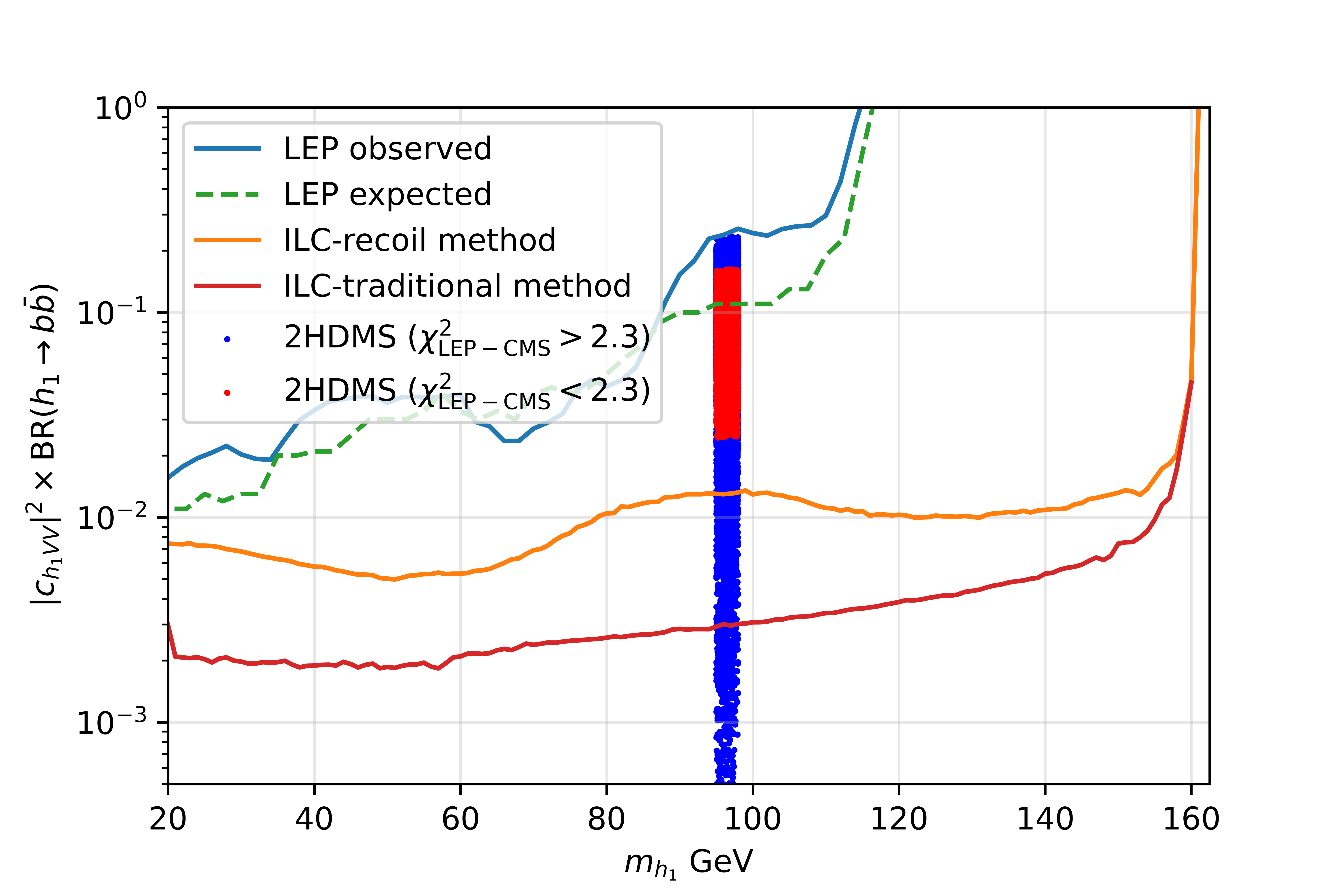}
    \caption{The plane of $m_{h_1}$ and $S_{95}$, which is defined as $\sigma(e^+ e^- \rightarrow Z h_1)/\sigma_{\text{SM}} \times \text{BR}(h_1\rightarrow b\bar{b})$. 
    The green dashed (blue) line indicate the expected (observed) limits at LEP~\cite{Barate:2003sz}. The orange and the red line show the reach of the ILC using the "recoil method" or the "traditional method" (see text). The red (blue) points indicate the parameter points within (outside of) the 1$\sigma$ range of the 96 GeV excesses.}
    \label{fig:detectlim}
\end{figure}

In \reffi{fig:detectlim} we show the plane of $m_{h_1}$ and the quantity $|c_{h_1VV}|^2 \times \br(h_1 \to b \bar b)$. The green dashed (blue) line indicate the expected (observed) limits at LEP~\cite{Barate:2003sz}, where the $2\,\sig$ excess at $\sim 96 \gev$ can be observed. The orange and the red line show the reach of the ILC using the "recoil method"~\cite{OPAL:2002ifx} or the "traditional method", see \citere{Drechsel:2018mgd} for details (and \citere{Wang:2018fcw} for a corresponding experimental analysis). This analysis assumed $\sqrt{s} = 250 \gev$ and an integrated luminosity of 500~fb$^{-1}$. The colored dots indicate the results from our parameter scan in the 2HDMS. The red (blue) points correspond to the parameter points inside (outside) the $1\,\sig$ ellipse of $\mu_{\text{CMS}}$ and $\mu_{\text{LEP}}$. One can observe that the red points, i.e.\ the ones describing the two excesses, are all well above the orange line. This shows that such light Higgs boson could be produced abundantly at the ILC.
The same conclusion holds for the N2HDM.

\begin{figure}[!]
    \centering
    \begin{subfigure}{.48\textwidth}
    \includegraphics[width=\textwidth]{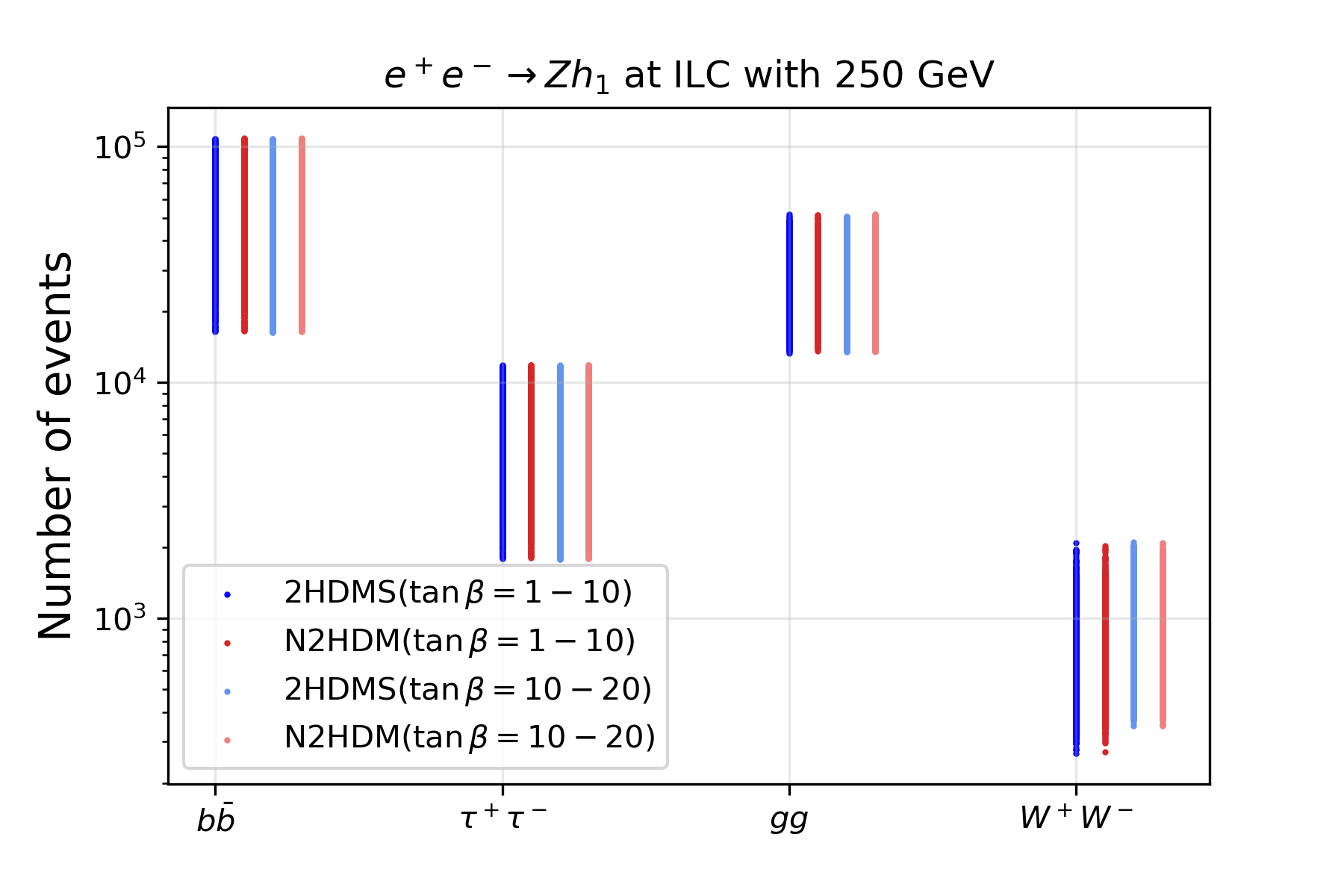}
    \caption{}
    \end{subfigure}
    \begin{subfigure}{.48\textwidth}
    \includegraphics[width=\textwidth]{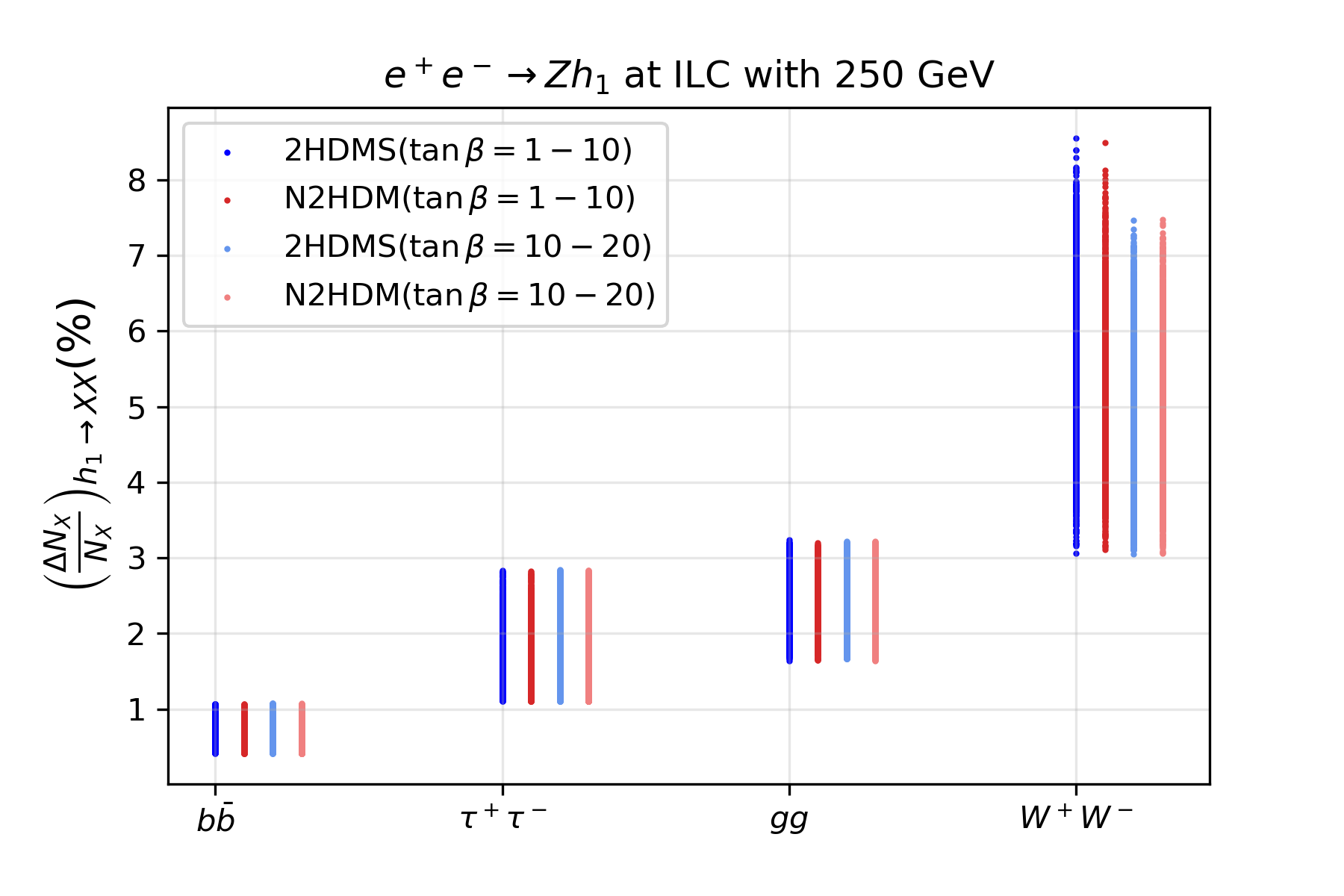}
    \caption{}
    \end{subfigure}
    \caption{Number of events (left) at the $h_1\rightarrow b\bar{b}$, $h_1\rightarrow \tau^+\tau^-$, $h_1\rightarrow gg$ and $h_1\rightarrow W^+ W^-$ final states produced by the Higgs-strahlung process at the ILC, and the respective uncertainties (right) for the 2HDMS and the N2HDM scan points, which are within the 1$\sigma$ ellipse of the 96 GeV excesses. The ILC center-of-mass energy is $\sqrt{s}=250 \gev$ and the integrated luminosity is $2\,\iab$.} 
    \label{fig:h1num}
\end{figure}

\begin{figure}[!]
    \centering
    \begin{subfigure}{.48\textwidth}
    \includegraphics[width=\textwidth]{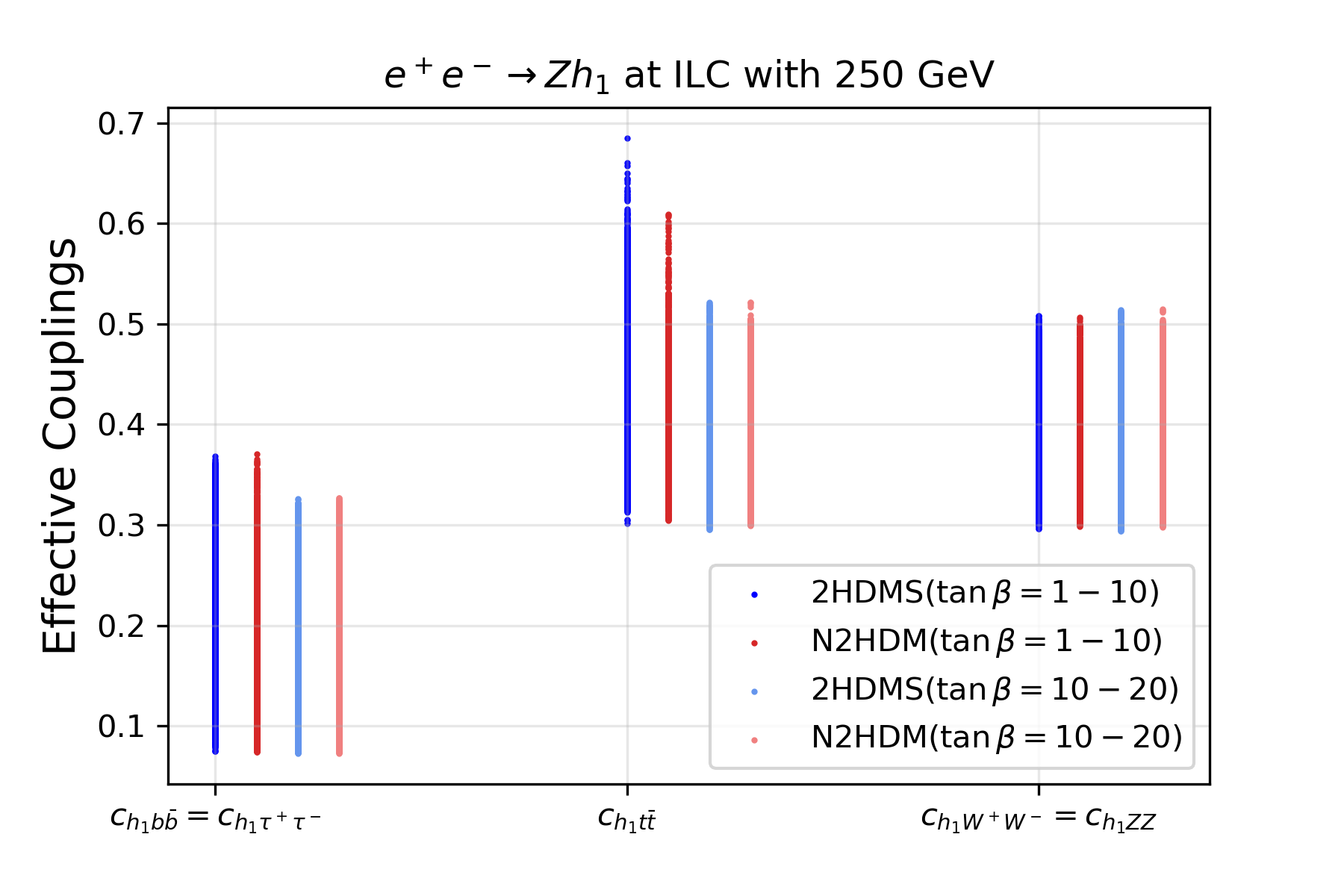}
    \caption{}
    \end{subfigure}
    \begin{subfigure}{.48\textwidth}
    \includegraphics[width=\textwidth]{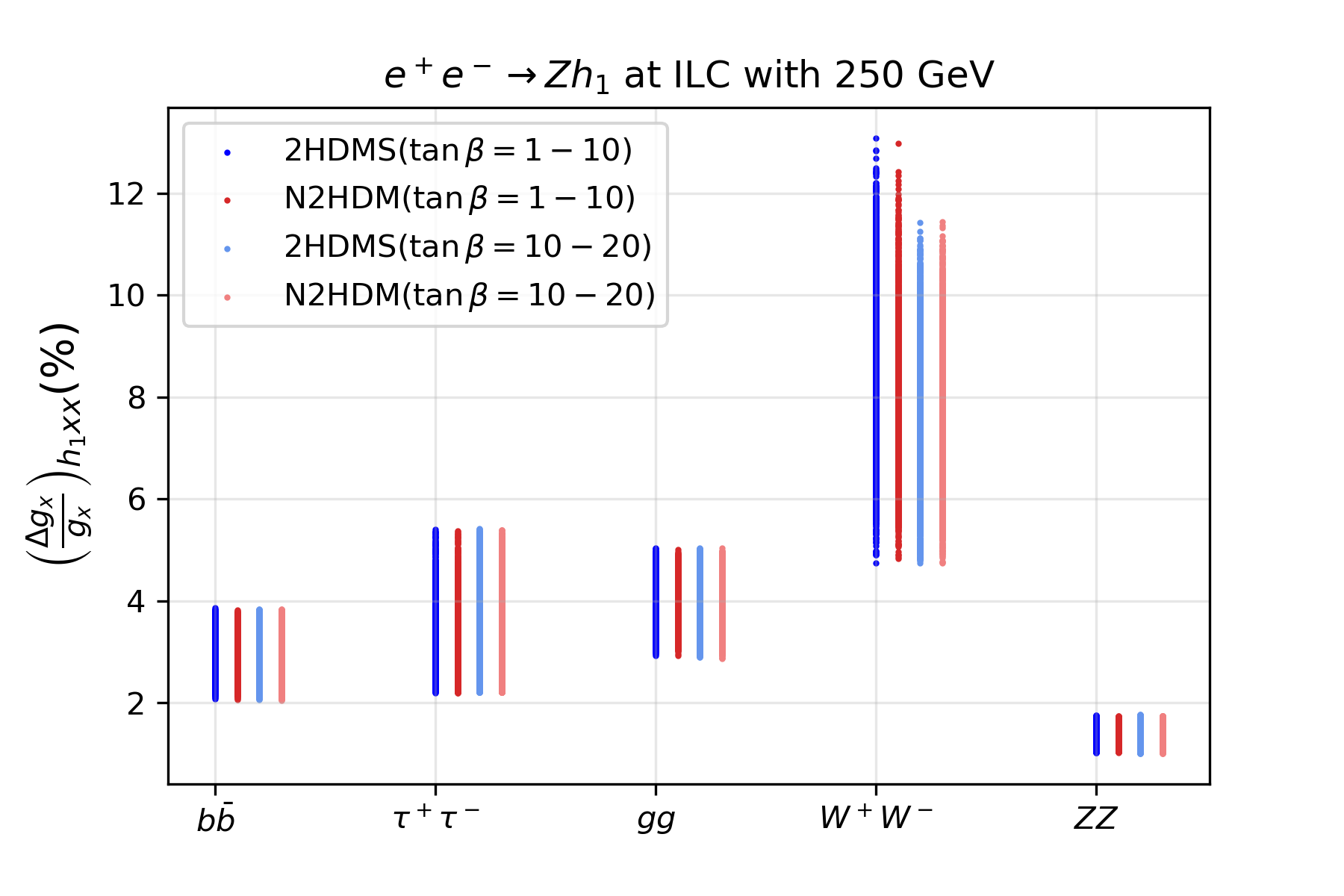}
    \caption{}
    \end{subfigure}
    \caption{Left: the effective couplings of the $h_1$ for both 2HDMS and N2HDM in the two $\tan\beta$ regions. Right: the anticipated coupling measurement uncertainties (see text). 
    }
\label{fig:h1numunc}
\end{figure}

In a second step we analyze the anticipated precision of the $h_1$~measurements that can be performed at the ILC, where we assume a center-of-mass energy of $\sqrt{s} = 250 \gev$ and an integrated Luminosity of $2 \,\iab$. We concentrate on the points within the $1\,\sig$ ellipse of the $96 \gev$ excesses, i.e.\ in the $\mu_{\text{CMS}}$-$\mu_{\text{LEP}}$ plane. In \reffi{fig:h1num} (left) we show the numbers of $h_1$ events produced in the Higgs-strahlung channel for the dominant decay modes. We directly compare the results for the 2HDMS and the N2HDM for the low and the high $\tb$ region. It can be seen that no relevant differences can be observed, neither between the two models, nor for the two $\tb$ regions. It is remarkable that, depending on the channel between $\sim 10^3$ and up to $10^5$ events can be expected. The statistical uncertainty for these numbers is shown in the right plot of \reffi{fig:h1num} (details can be found in App.~\ref{sec:coup-eval}). 

In the left plot of \reffi{fig:h1numunc} we show the predictions for the effective couplings, which are the same for $c_{h_1bb}$ and $c_{h_1\tau\tau}$, as well as for $c_{h_1ZZ}$ and $c_{h_1WW}$. The only visible difference between the low and high $\tb$ region is the somewhat enlarged range of $c_{h_1tt}$, which is found in the low $\tb$ region. Naively, one would expect a corresponding enhancement in the number of $gg$ events in the left plot of \reffi{fig:h1num}. However, the corresponding branching ratio is largely driven by the decay $h_1 \to b \bar b$, and no direct correspondence of $c_{h_1tt}$ and $\br(h_1 \to gg)$ is found (see also the numbers for the best-fit points in \reftas{tab:bestfit} and \ref{tab:bestfit2}). Finally in the right plot of \reffi{fig:h1numunc} we show the anticipated precision for the $h_1$ coupling measurement at the ILC (details about this evaluation are given in App.~\ref{sec:coup-eval}). The coupling of the $h_1$ to $b\bar b$, $\tau^+\tau^-$, $gg$, and $W^+W^-$ are determined from the respective decays, whereas the coupling to $ZZ$ is determined from the Higgs-strahlung production. It is expected that the coupling of the $h_1$ to $b\bar b$ can be measured with an uncertainty between $2\,\%$ and $\sim 3.5\,\%$. For $\tau^+\tau^-$ and $gg$, the precision is expected to be only slightly worse. Because of the smaller coupling to $W$~bosons, the corresponding uncertainty is found between $\sim 4.5\,\%$ and $\sim 12\,\%$. The highest precision, however, is expected from the light Higgs-boson production via radiation from a $Z$~boson, where an accuracy between $1\,\%$ and $2\,\%$ is anticipated. While these precisions are the same for the two models under investigation and as well as for the two $\tb$ regions, they will nevertheless allow for a high-precision test of the 2HDMS/N2HDM predictions.


\subsection{Measurement of the \boldmath{$h_2$} couplings}
\label{sec:ilc-h125}

\begin{figure}[!]
    \centering
    \begin{subfigure}{.48\textwidth}
        \includegraphics[width=\textwidth]{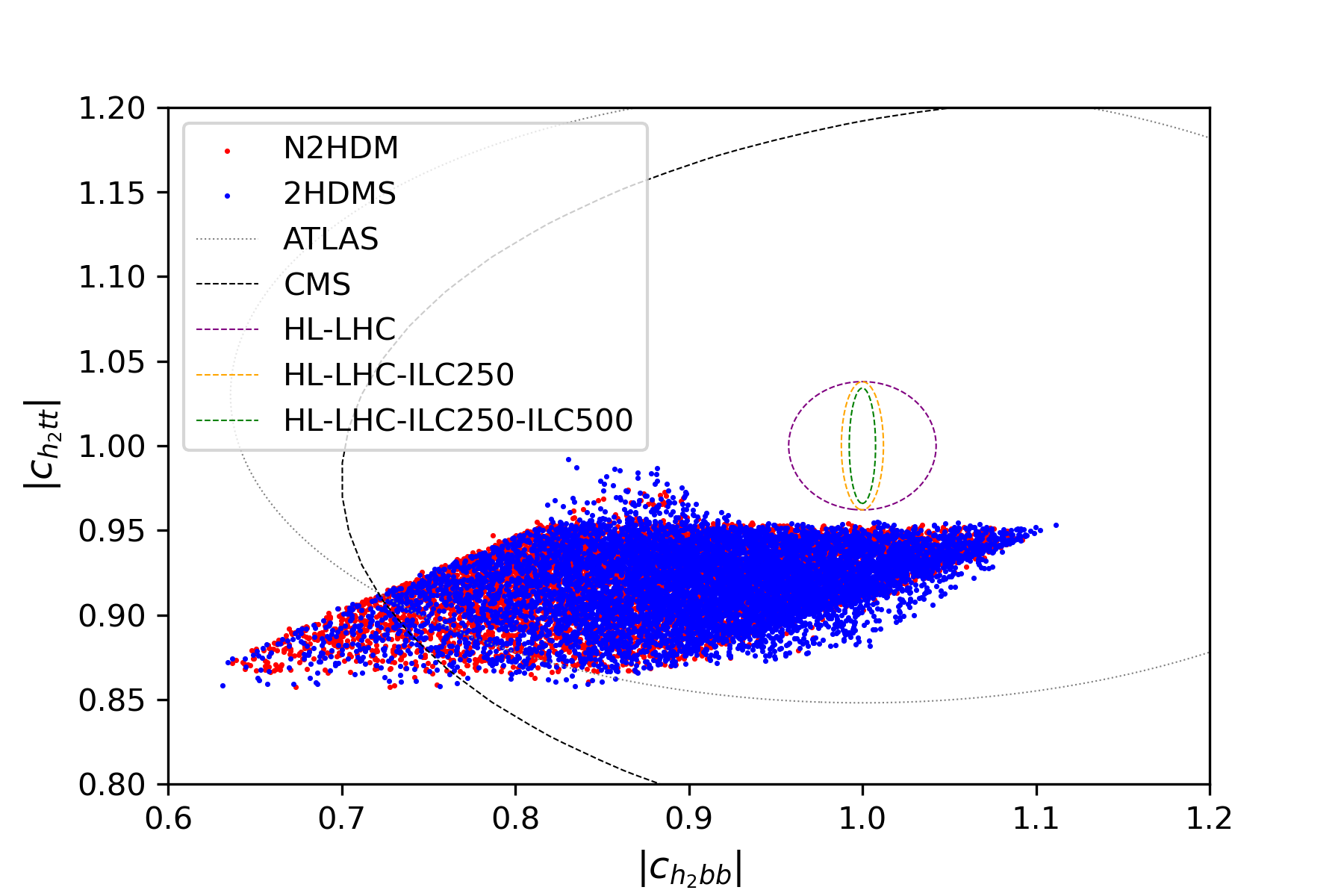}
     \caption{}
    \label{fig:cph2bbtt}
    \end{subfigure}
    \begin{subfigure}{.48\textwidth}
        \includegraphics[width=\textwidth]{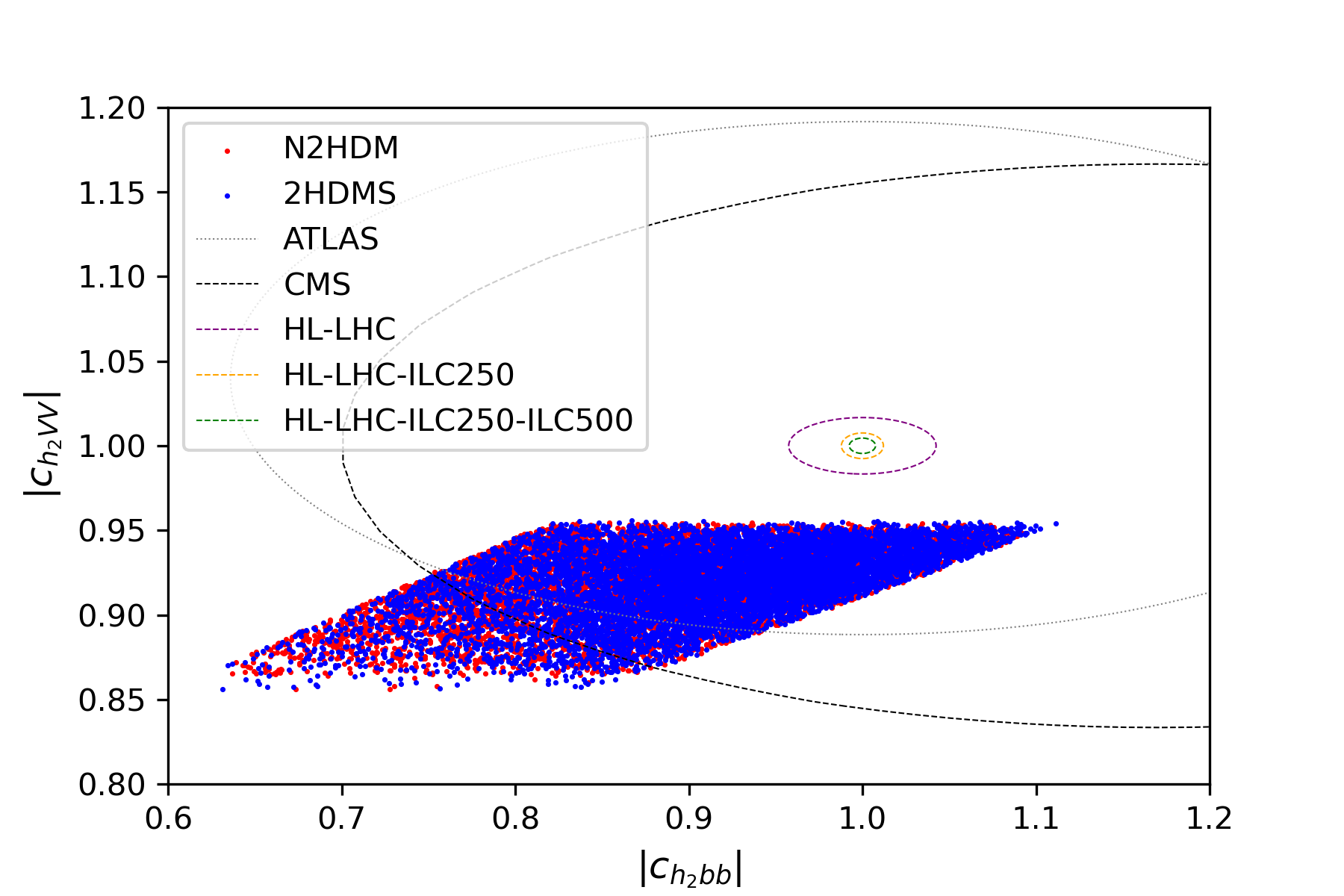}
     \caption{}
    \label{fig:cph2bbvv}
    \end{subfigure}
\begin{subfigure}{.48\textwidth}
        \includegraphics[width=\textwidth]{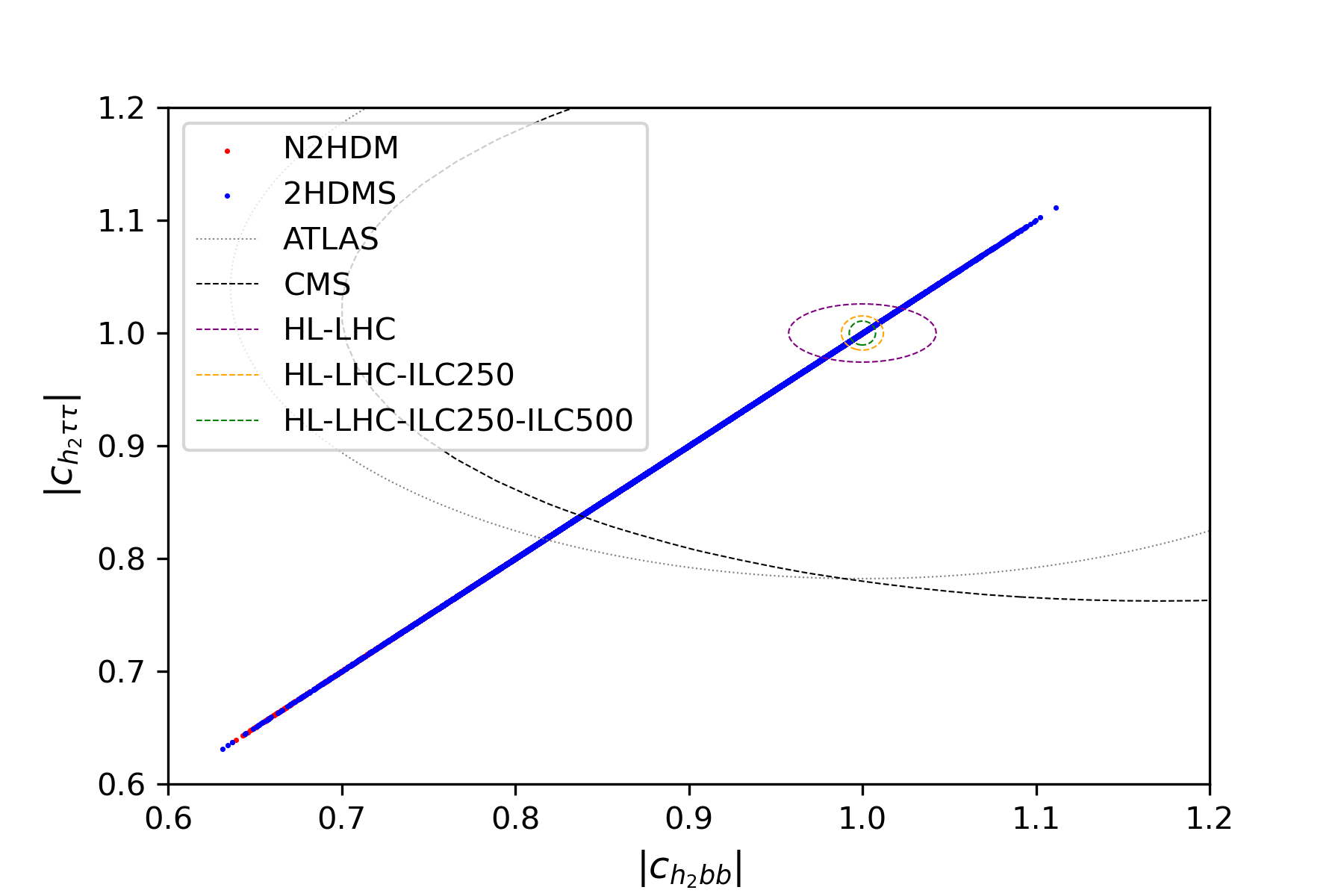}
     \caption{}
     \label{fig:cph2bbtata}
\end{subfigure}
      \caption{The effective couplings of the SM-like Higgs-boson $h_2$. Shown are $|c_{h_2}bb$ vs.\ $c_{h_2tt}$ (upper left), $c_{h_2VV}$ (upper right) and $c_{h_2\tau\tau}$ (lower plot). The blue (red) points show the 2HDMS (N2HDM) points within the $1\,\sig$ range of the $96 \gev$ excesses, with the two $\tb$ ranges combined. The black dotted (dashed) ellipse indicate the current ATLAS (CMS) $1\,\sig$ limits. The HL-LHC expectation, centered around the SM value, are shown as dashed violet ellipse. The orange (green) dashed ellipses indicate the improvements expected from the ILC at $250 \gev$ (additionally at $500 \gev$).}
      \label{fig:cph2}
\end{figure}

The Higgs boson observed at $\sim 125 \gev$ at the LHC can also serve for the exploration of BSM models. The extended Higgs-boson sector of the 2HDMS/N2HDM, in particular the mixing of the lighter doublet with the singlet, yields deviations of the $h_2$ couplings from their SM expectations. In \reffi{fig:cph2} we compare the predictions of the 2HDMS (blue points) and the N2HDM (red points) for the effective $h_2$ couplings with the experimental accuracies. Only points within the $1\,\sig$ ellipse of the $96 \gev$ excesses are used, where the two $\tb$ regions have been combined. Shown are $c_{h_2}bb$ vs.\ $c_{h_2tt}$ (upper left), $c_{h_2VV}$ (upper right) and $c_{h_2\tau\tau}$ (lower plot).  The black dotted (dashed) ellipse indicate the current ATLAS (CMS) $1\,\sig$ limits (see~\citeres{ATLAS:2018doi}and~\cite {CMS:2018uag}). The HL-LHC expectation~\cite{Cepeda:2019klc}, centered around the SM value, are shown as dashed violet ellipse. The orange (green) dashed ellipses indicate the improvements expected from the ILC at $250 \gev$ (additionally at $500 \gev$), based on \citere{Bambade:2019fyw}.
All the points are roughly within the $2\,\sigma$ range of the current Higgs-boson rate measurements at the LHC, because of the $\mathtt{HiggsSignals}$ constraint. No relevant difference between the two models can be observed. While $c_{h_2bb} = c_{h_2\tau\tau}$ can reach the SM value (which by definition of the effective couplings is~1), the couplings to top quarks and to gauge bosons always deviate at least $\sim 5\,\%$ from the SM prediction. We have checked explicitly that this is due to the agreement with the $96 \gev$ excesses. For the coupling to top quarks, depending which point in the parameter space is realized, possibly no deviation can be observed, neither with the HL-LHC, nor with the ILC precision. The situation is different for the $h_2$ coupling to gauge bosons. The HL-LHC precision might still yield a significance below the $\sim 3\,\sig$ level. The in this case strongly improved ILC precision, on the other hand, yields for all parameter points of the 2HDMS or the N2HDM a deviation from the SM prediction larger than $5\,\sig$. Consequently, the anticipated high-precision $h_2$ coupling measurements at the ILC will always either rule out the 2HDMS/N2HDM, or refute the SM prediction. On the other hand, no distinction between the two models will be visible via the $h_2$ coupling determinations.

\section{Conclusions and Outlook}
\label{sec:conclusions}

We analyzed a $\sim 3\,\sigma$ excess (local)
in the di-photon decay mode at $\sim 96 \gev$ as reported by
CMS, together with a $\sim 2\,\sigma$ excess (local) in the
$b \bar b$ final state at LEP in the same mass range.
These two excesses can be interpreted as a new Higgs boson in extended SM models. 
Specifically we investigate the 2HDM type~II with an additional singlet that 
can be either real (N2HDM) or complex (2HDMS). Besides the nature of the
singlet, the two models differ in their symmetry structure. Both obey
the \zet2 symmetry present already in the 2HDM to avoid flavour changing
neutral currents at the tree level. The N2HDM additionally obeys a second 
\zet2 symmetry, while the 2HDMS obeys a \zet3 symmetry such that 
this Higgs sector corresponds to the Higgs sectors of the NMSSM. 
In both models we have taken the lightest CP-even Higgs boson, $h_1$, at $96 \gev$,
while the second-lightest, $h_2$, was set to $125 \gev$.

All relevant constraints are taken into account in our analysis. 
The first set are theoretical constraints from perturbativity and the requirement that the
minimum of the Higgs potential is a stable global minimum. The second set are 
experimental constraints, where we take into account
the direct searches for additional Higgs bosons from LEP. the Tevatron
and the LHC, as well as the measurements of the properties of the Higgs
boson at $\sim 125 \gev$. Furthermore, we include bounds from flavor
physics and from electroweak precision data. 

We find that both models, the 2HDMS and the N2HDM can fit simultaneously the two excesses 
very well. (For a previous analysis of the N2HDM, see \citere{Biekotter:2019kde}.) 
Neither the different particle content, nor the different symmetry structures 
have a relevant impact on how well the two models describe the excesses. 

In a second step we have analyzed the couplings of the $h_2$ to SM fermions. 
In both models it was shown that in particular the coupling to gauge bosons differs 
substantially from its SM prediction. With the anticipated accuracy at the HL-LHC
a $\sim 3\,\sig$ effect would be visible. At a future $e^+e^-$ collider, where 
concretely we employed ILC predictions, a signal at $5\,\sig$ or more would be visible.

In the final step of our analysis (also going beyond \citere{Biekotter:2019kde}) we 
analyzed with which precision the couplings of the $96 \gev$ Higgs boson could be
measured at the ILC. The highest precision at the level of $1-2\,\%$ is expected for 
the $h_1ZZ$ coupling, which is determined from the $h_1$ production in the 
Higgs-strahlung channel. Other coupling precisions are expected at the level of 
$2-12\,\%$. 

In all future coupling determinations no difference between the N2HDM and the 2HDMS
was found. Consequently, a distinction between the two models can be found only in 
the different particle content, or in the symmetry structure. The previous could manifest
itself in the properties of the lighter CP-odd Higgs boson, which is always doublet-like 
in the N2HDM, but can be singlet-like in the 2HDMS. The different symmetry structure
has an impact on trilinear Higgs couplings. It could manifest itself concretely
(in the model realizations we have analyzed) in the decays of the heaviest CP-even Higgs
boson. We leave both kinds of analyses for future work.

\subsection*{Acknowledgements}

We thank M.~Cepeda, J.~Tian and C.~Schappacher for invaluable help 
in the light Higgs-boson coupling analysis at the ILC.
We thank T.~Bietk\"otter
and
J.~Wittbrodt 
for helpful discussions.
C.L., G.M.-P.\ and S.P.\ acknowledge support by the Deutsche Forschungsgemeinschaft (DFG, German Research Foundation) under Germany’s Excellence Strategy – EXC 2121 "Quantum Universe" – 390833306.
The work of S.H.\ is supported in part by the
MEINCOP Spain under contract PID2019-110058GB-C21 and in part by
the AEI through the grant IFT Centro de Excelencia Severo Ochoa SEV-2016-0597.



\begin{appendices}
\section*{Appendix}
\section{Tree level Higgs masses}
\label{appendix:a}

In the following we discuss and compare the tree-level mixing matrices and physical basis in the 2HDMS and the N2HDM.

\subsection{The 2HDMS}

By taking the second derivative of the scalar potential, one can obtain the tree level Higgs mass matrices:
\begin{equation}
    M_{Sij}^2=\frac{\partial^2V}{\partial\rho_i\partial\rho_j}\Big|_{\substack{\Phi_1=v_1\\\Phi_2=v_2\\S=v_S}},\qquad M_{Pij}^2=\frac{\partial^2V}{\partial\eta_i\partial\eta_j}\Big|_{\substack{\Phi_1=v_1\\\Phi_2=v_2\\S=v_S}},\qquad
    M_{Cij}^2=\frac{\partial^2V}{\partial\chi_i\partial\chi_j}\Big|_{\substack{\Phi_1=v_1\\\Phi_2=v_2\\S=v_S}}
\end{equation}
For the CP-even Higgs mass matrix one finds, 
\begin{subequations}
\begin{align}
M_{S11}^2&=2\lambda_1v^2\cos^2\beta+(m_{12}^2-\mu_{12}v_S)\tan\beta~,\\
M_{S22}^2&=2\lambda_2v^2\sin^2\beta+(m_{12}^2-\mu_{12}v_S)\cot\beta~,\\
M_{S12}^2&=(\lambda_{3}+\lambda_4)v^2\sin2\beta-(m_{12}^2-\mu_{12}v_S)~,\\
M_{S13}^2&=(2\lambda'_1v_S\cos\beta+\mu_{12}\sin\beta)v~,\\
M_{S23}^2&=(2\lambda'_2v_S\sin\beta+\mu_{12}\cos\beta)v~,\\
M_{S33}^2&=\frac{\mu_{s1}}{2}v_S+\lambda''_3v_S^2-\mu_{12}\frac{v^2}{2v_S}\sin2\beta~.
\label{eq:masscpeven}
\end{align}
\end{subequations}
The mass matrix of CP-odd Higgs-sector is given by, 
\begin{subequations}
	\begin{align}
	M_{P11}^2&=(m_{12}^2-\mu_{12}v_S)\tan\beta~,\\
	M_{P22}^2&=(m_{12}^2-\mu_{12}v_S)\cot\beta~,\\
	M_{P12}^2&=-(m_{12}^2-\mu_{12}v_S)~,\\
	M_{P13}^2&=\mu_{12}v\sin\beta~,\\
	M_{P23}^2&=-\mu_{12}v\cos\beta~,\\
	M_{P33}^2&=-\frac{3}{2}\mu_{S1}v_S-\mu_{12}\frac{v^2}{2v_S}\sin2\beta~. 
	\label{eq:masscpodd}
	\end{align}
\end{subequations}
The charged Higgs-boson mass is obtained as, 
\begin{equation}
M_C^2=2(m_{12}^2-\mu_{12}v_S)\csc2\beta-\lambda_4v^2~.
\end{equation}

Since we parameterize the CP-even mixing matrix as the $3\times3$ rotation matrix $R$, see \refeq{eq:rot}, one can perform the diagonalization and express each mass-matrix element in terms of the mixing elements and the eigenvalues. In the CP-even Higgs-boson sector one has, 
\begin{equation}
M^2_{Sij}=\sum_{n=1}^3 m^2_{h_n}R_{ni}R_{nj}~.
\end{equation}
One can express the input parameters in terms of the Higgs-boson masses and mixing angles. Therefore we have the following relations to convert the input parameters in the original Lagrangian to the input parameters in the mass basis, 
\begin{align}
&\mu_{12}=\frac{m_{a_2}^2-m_{a_1}^2}{v}\sin\alpha_4\cos\alpha_4\label{eq:mu12-a4}~,\\
&v_S=\frac{m^2_{12}-\tilde{\mu}^2\sin\beta\cos\beta}{\mu_{12}}~,\\
&\mu_{S1}=-\frac{2}{3v_S}\left(\sin^2\alpha_4m_{a_1}^2+\cos^2\alpha_4m_{a_2}^2+\frac{v^2}{2v_S}\sin2\beta\mu_{12} \right)~,\\
&\lambda_1=\frac{1}{2v^2\cos^2\beta}\left(\sum_im^2_{h_i}R^2_{i1}-\tilde{\mu}^2\sin^2\beta\right)~,\\
&\lambda_2=\frac{1}{2v^2\sin^2\beta}\left(\sum_im^2_{h_i}R^2_{i2}-\tilde{\mu}^2\cos^2\beta\right)~,\\
&\lambda_3=\frac{1}{v^2}\left(\frac{1}{\sin2\beta}\sum_im^2_{h_i}R_{i1}R_{i2}+m^2_{h^\pm}-\frac{\tilde{\mu}^2}{2}\right)~,\\
&\lambda_4=\frac{\tilde{\mu}^2-m^2_{h^\pm}}{v^2}~,\\
&\lambda'_1=\frac{1}{2v_Sv\cos\beta}\left(\sum_i m^2_{h_i}R_{i1}R_{i3}-\mu_{12}v\sin\beta \right)~,\label{eq:lam1p}\\
&\lambda'_2=\frac{1}{2v_Sv\sin\beta}\left(\sum_i m^2_{h_i}R_{i2}R_{i3}-\mu_{12}v\cos\beta \right)~,\\
&\lambda''_3=\frac{1}{v_S^2}\left(\sum_im^2_{h_i}R^2_{i3}+\mu_{12}\frac{v^2}{2v_S}\sin2\beta-\frac{\mu_{S1}}{2}v_S\right)~,
\end{align}
where we define the parameter $\tilde{\mu}^2$ as, 
\begin{equation}
\tilde{\mu}^2=\frac{m^2_{12}-v_S\mu_{12}}{\sin\beta\cos\beta}
\equiv \cos^2\alpha_4m_{a_1}^2+\sin^2\alpha_4m_{a_2}^2~.
\end{equation}


\subsection{The N2HDM}

The symmetric CP-even Higgs mass matrix in the N2HDM is obtained via a diagonalization with the same Rotation matrix $R$ in \refeq{eq:rot},
\begin{subequations}
\begin{align}
M_{S11}^2&=2\lambda_1v^2\cos^2\beta+m_{12}^2\tan\beta~,\\
M_{S22}^2&=2\lambda_2v^2\sin^2\beta+m_{12}^2\cot\beta~,\\
M_{S12}^2&=(\lambda_{3}+\lambda_4+\lambda_5)v^2\sin\beta\cos\beta-m_{12}^2~,\\
M_{S13}^2&=2\lambda_7v_S\cos\beta v~,\\
M_{S23}^2&=2\lambda_8v_S\sin\beta v~.\\
M_{S33}^2&=\lambda_6 v_S~.
	\end{align}
\end{subequations}

In the CP-odd sector the difference between the two models is characterized by the abscence of a CP-odd mixing matrix due to the missing additional pseudoscalar-like Higgs in the N2HDM. In the scalar sector we have the same form as for the 2HDMS. It sould be noted, however, that in the 2HDMS the appearance of the parameters $\mu_{12}$ and $\mu_{S1}$ leads to additional contributions. 

One can perform a change of basis from the parameters of the potential to the physical masses and mixing angles similar to the 2HDMS. One finds, 
\begin{align}
&\lambda_1=\frac{1}{v^2\cos^2\beta}\left(\sum_im^2_{h_i}R^2_{i1}-\hat{\mu}^2\sin^2\beta\right)~,\\
&\lambda_2=\frac{1}{v^2\sin^2\beta}\left(\sum_im^2_{h_i}R^2_{i2}-\hat{\mu}^2\cos^2\beta\right)~,\\
&\lambda_3=\frac{1}{v^2}\left(\frac{1}{\sin\beta\cos\beta}\sum_im^2_{h_i}R_{i1}R_{i2}+2m^2_{h^\pm}-\hat{\mu}^2\right)~,\\
&\lambda_4=\frac{1}{v^2}\left(\hat{\mu}^2+m_A^2-2m^2_{h^\pm}\right)~,\\
&\lambda_5=\frac{1}{v^2}\left(\hat{\mu}^2-m_A^2\right)~,\\
&\lambda_6=\frac{1}{v_S^2}\sum_i m^2_{h_i}R_{i3}~,\\
&\lambda_7=\frac{1}{v v_S \cos\beta}\left(\sum_i m^2_{h_i}R_{i1}R_{i3}\right)~,\\
&\lambda_8=\frac{1}{v v_S \sin\beta}\left(\sum_i m^2_{h_i}R_{i2}R_{i3}\right)~,\\
\end{align}
where $\hat{\mu}^2$ is here defined as, 
\begin{equation}
\hat{\mu}^2=\frac{m^2_{12}}{\sin\beta\cos\beta}~.
\end{equation}

\section{Evaluation of experimental coupling uncertainties for a light Higgs boson}
\label{sec:coup-eval}

In this section we describe in detail how we estimate the experimental
uncertainties of the coupling measurements of a Higgs boson below
$125 \gev$ based on ILC250 measurements.%
\footnote{We thank M.~Cepeda for invaluable help in this section. We also thank J.~Tian for providing ILC numbers.}


\subsection{SM Higgs-boson results}

In this subsection we denote the SM Higgs boson as $h$ and assume a mass of
$125 \gev$.
The cross section at the ILC250 is given as
\begin{align}
  \sig(e^+e^- \to Zh) = 206 \,\fb~.
  \label{sig-sm}
\end{align}
The BRs are given taken from~\citere{yr4} and summarized in \refta{tab:br-sm}.

\begin{table}[h]
\centering
\begin{tabular}{l|rrrrr} \hline
  final state & $b\bar b$ & $c \bar c$ & $gg$ & $\tau^+\tau^-$ & $WW^{*}$ \\
    \hline
    BR & 0.582 & 0.029 & 0.082 & 0.063 & 0.214 \\
    \hline
\end{tabular}
\caption{BRs of the SM Higgs boson~\cite{yr4}.}
\label{tab:br-sm}
\end{table}

The SH Higgs coupling uncertainties have been obtained in
\citere{Barklow:2017suo} (Tab. 2), assuming $\cL_{\rm int} = 2 \iab$ at
$\sqrt{s} = 250 \gev$ (i.e.\ the ILC250). The results are given  in \refta{tab:gx-sm}.

\begin{table}[h]
\centering
\begin{tabular}{l|rrrrrr} \hline
  coupling & $b\bar b$ & $c \bar c$ & $gg$ & $\tau^+\tau^-$ & $WW$ & $ZZ$ \\
    \hline
  rel. unc. [\%] & 1.04 & 1.79 & 1.60 & 1.16 & 0.65 & 0.66 \\
    \hline
\end{tabular}
\caption{Relative uncertainties in the SM Higgs couplings,
  $\De\gx/\gx$, at the ILC250~\cite{Barklow:2017suo}. }
\label{tab:gx-sm}
\end{table}

The numbers for the ratio of signal over background events, 
$S/B (=: f_h)$ of a SM Higgs boson at $125 \gev$
at the ILC250 are given in
\refta{tab:signalbackground}~\cite{snowmass13}.
The $hZZ$ coupling is determined directly from the cross section, where
the $q \bar q h$ mode can be neglected. The other couplings should be
taken in the $q \bar q h$ mode, corresponding effectively to
$e^+e^- \to Z^* \to Z h \to q \bar q h$ (with the subsequent Higgs decay).

\begin{table}[h]
\centering
\begin{tabular}{l|cc} \hline
  measurement & efficiency & $S/B$ \\
    \hline
    $\sig_{Zh}$ in $\mu^+\mu^-h$ & 88\% & 1/1.3 \\
    $\sig_{Zh}$ in $e^+e^-h$     & 68\% & 1/2.0 \\
    $\br(h \to b \bar b)$     in $q \bar q h$ & 33\% & 1/0.89 \\
    $\br(h \to c \bar c)$     in $q \bar q h$ & 26\% & 1/4.7 \\
    $\br(h \to gg)$           in $q \bar q h$ & 26\% & 1/13 \\
    $\br(h \to \tau^+\tau^-)$ in $q \bar q h$ & 37\% & 1/0.44 \\
    $\br(h \to WW)$           in $q \bar q h$ & 2.6\% & 1/0.96 \\
    \hline
\end{tabular}
\caption{Numbers for S/B at the ILC250~\cite{snowmass13}.}
\label{tab:signalbackground}
\end{table}


\subsection{Basic signal-background statistics}
\label{sec:stat}

In this section we use the following notation: 
\ns $(\equiv S)$: number of signal events; 
\nb $(\equiv B)$: number of background events;
\nt: total number of events with. Then one finds, 
\begin{align}
  \ns &= \nt - \nb~,\\
  f &= \ns/\nb~.
\end{align}
The background is taken after cuts, i.e.\ ``irreducable background'',
likely to be small at an $e^+e^-$ collider.
For the uncertainties we have
\begin{align}
  \De \ns^2 &= \De \nt^2 + \De \nb^2~.
\end{align}
The uncertainty of the total number of events scales like
\begin{align}
  \De\nt &= \sqrt{\nt}~.
\end{align}
The uncertainty of the background goes like
\begin{align}
  \De\nb &= \esb \cdot \nb~,
\end{align}
where $\esb$ denotes the relative uncertainty for background estimation
(which cancels out later).
Therefore,
\begin{align}
  \De\ns^2 &= \LP \sqrt{\nt} \RP^2 + \LP \esb \cdot \nb \RP^2~,\\
  \De\ns &= \sqrt{\LP \ns + \nb \RP + \esb^2 \nb^2}~.
\end{align}
If the background is known perfectly, one has overall uncertainty fully
dominated by the purely statistical uncertainty, 
\begin{align}
  \esb &= 0~,\\
  \De\ns &= \sqrt{\ns + \nb} \non \\
    &= \sqrt{\ns + \ns/f} \non \\
    &= \sqrt{\ns \LP 1 + 1/f \RP} \non \\
    &= \sqrt{\ns} \cdot \sqrt{1 + 1/f}~,\\
    \De\ns/\ns &= \frac{1}{\sqrt{\ns}} \cdot \sqrt{1 + 1/f}~.
\label{eq:dns}
\end{align}
Consequently, the uncertainty improves with $\sqrt{\ns}$ if
$f = \ns/\nb \gg 1$. On the other hand, 
if $f$ is small, one wins less from the gain in statistics.


\subsection{Evaluation of uncertainties in the Higgs couplings}
\label{sec:evalunc}

\subsubsection{Cross section evaluation}

\begin{figure}[h]
\begin{center}
\includegraphics[width=0.65\textwidth]{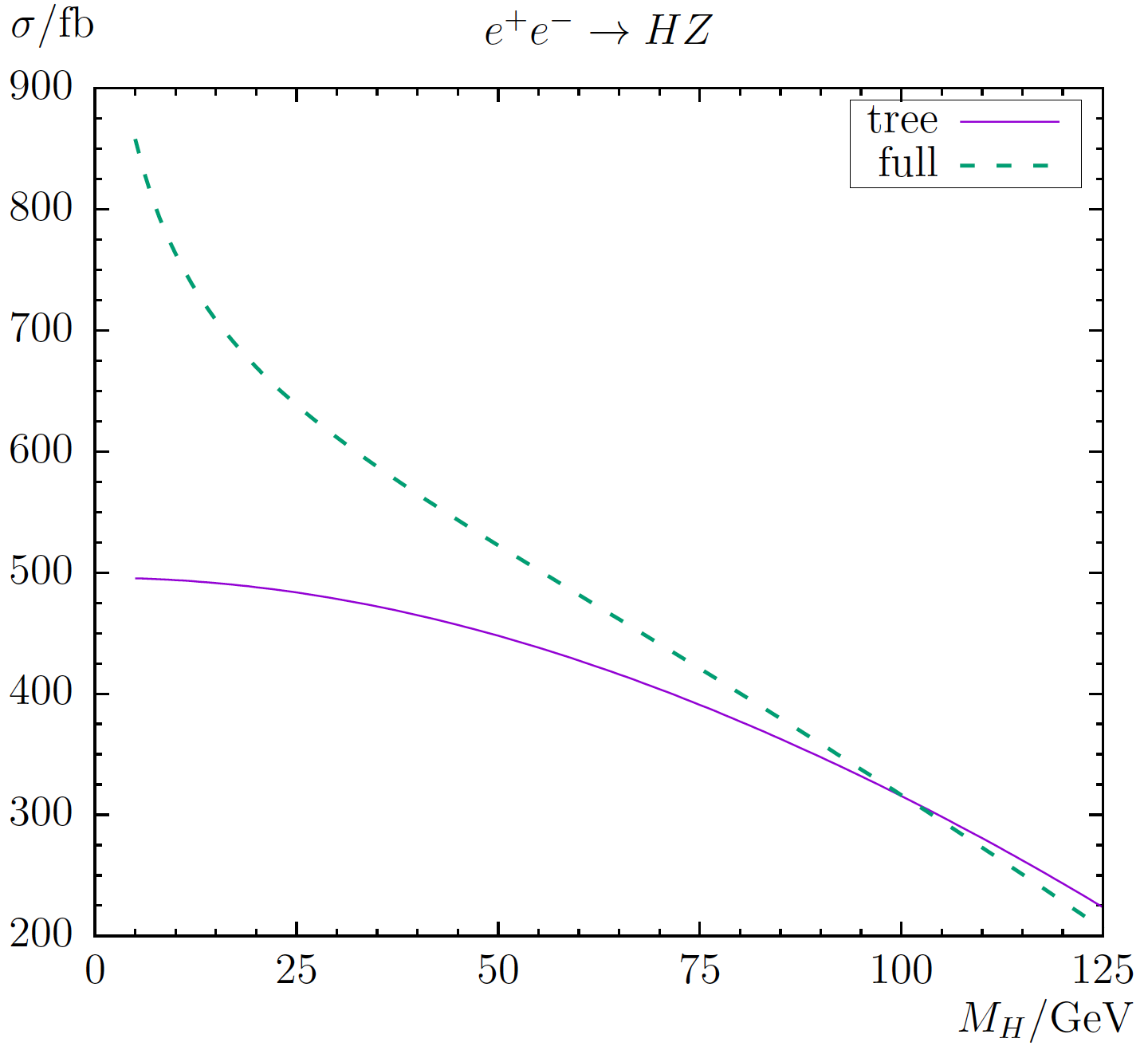}
\caption{\label{fig:eeHZ}
  Production cross section for ``SM Higgs bosons'' with
  $\MH \le 125 \gev$~\cite{eeHZ}.
}
\end{center}
\end{figure}

The production cross section for a Higgs $\phi$ at an $e^+e^-$
collider is evaluated as
\begin{align}
  \sig(e^+e^- \to \phi Z) =
  \sig_{\rm SM}(e^+e^- \to H_{\rm SM}^\phi Z) \times |c_{\phi VV}|^2 ~.
\end{align}
Here $H_{\rm SM}^\phi$ is the SM Higgs boson with a hypothetical
mass equal to $m_\phi$. $c_{\phi VV}$ is the coupling strength
of the $\phi$ to two gauge bosons ($V = W^\pm, Z$) relative to the SM
value. In \reffi{fig:eeHZ} we show the evaluation of
$\sig_{\rm SM}(e^+e^- \to H_{\rm SM}^\phi Z)$ (where $H_{\rm SM}^\phi$
is labeled~$H$) at the tree-level (``tree'') and the full one-loop
level (``full'')~\cite{eeHZ}, including soft and hard QED radiation.%
\footnote{We thank C.~Schappacher for providing the calculation.}%
~One can see that the loop corrections are important for the reliable evaluation
of this cross section. Explicit numbers are given in \refta{tab:eeHZ}.
Multiplying the loop corrected cross section with $|c_{\phi VV}|^2$ is
an approximation that works well for $m_\phi \gsim 75 \gev$ and requires
more scrutiny for the lowest Higgs-boson masses.

\begin{table}[h]
\centering
\begin{tabular}{l|rrrrrrrrrrrrrr} \hline
  $m_\phi$ [GeV] &
     5 &  10 &  20 &  30 &  40 &  50 &  60 &  70 &  80 &  90 &  96 & 100
 & 110 & 120 \\ 
    \hline
    $\sig_{HZ}$ [fb] &
   858 & 763 & 670 & 611 & 565 & 523 & 482 & 441 & 400 & 359 & 333 & 316
 & 273 & 228 \\
    \hline
\end{tabular}
\caption{Production cross section for ``SM Higgs bosons'' with
  $m_\phi \equiv \MH \le 125 \gev$~\cite{eeHZ}.}
\label{tab:eeHZ}
\end{table}


\subsubsection{Signal over background for the new Higgs boson}

An important element for the evaluation of the Higgs-boson coupling
uncertainties is the number of signal over background events for~$\phi$,
\begin{align}
  \LP \frac{\ns}{\nb} \RP_\phi =: f_\phi
  \label{fphi}
\end{align}
relative to the SM value(s) as given in \refta{tab:signalbackground},
\begin{align}
  \LP \frac{\ns}{\nb} \RP_h =: f_h~
  \label{fh}
\end{align}
with
\begin{align}
  \LP \frac{\ns}{\nb} \RP_h / \LP \frac{\ns}{\nb} \RP_\phi = f_h/f_\phi =: D~.
  \label{fphifh}
\end{align}
Unfortunately, there is no (neither general nor model specific)
evaluation of $f_\phi$ or $D$ available. However, for the ILC500
``SM-like'' Higgs bosons with masses above and below $125 \gev$ have
been simulated~\cite{jenny}. One finds that in this case very roughly $D \approx 2$ 
can be assumed. For our evaluation we use $D = 3$ as a conservative value.



\subsubsection{Relating signal events to Higgs couplings}

In this subsection we derive the evaluation of the uncertainties in the (light) Higgs-boson couplings. We denote the
generic coupling of a Higgs $\phi$ to another particle~$x$ as \gx.
There are two cases:
\begin{itemize}

\item[(i)]
The coupling is determined via the decay $\phi \to xx$. The number of
signal events is given by
  
\begin{align}
\label{eq:ns}
  \ns &= \cL_{\rm int} \times \sig(e^+e^- \to \phi Z)
  \times \br(\phi \to xx) \times \epsilon_{\rm sel}
  \times \br(Z \to q \bar q)~, \\
  \br(\phi \to xx) &= \frac{\Ga(\phi \to xx)}{\Gatot}~,
\end{align}
where $\epsilon_{\rm sel}$ is the selection efficiency. In the formulas
below $\epsilon_{\rm sel}$ and $\br(Z \to q \bar q)$ cancel out,
but they enter in the $\ns/\nb$ evaluation, i.e.\ in the numbers of
$S/B$ given in \refta{tab:signalbackground}, as well as in the coupling
precisions given in \refta{tab:gx-sm}. 
The decay channel $\phi \to xx$ gives not only $\Ga(\phi \to xx)$, but
also contributes to \Gatot. For simplicity we assume
\begin{align}
  \br(\phi \to xx) &= \frac{\gx^2}{\gx^2 + g^2}~,
\end{align}
where $g^2$ (modulo canceled prefactors) summarizes the other
contributions. The relative strength between \gx\ and $g$ is given by
\begin{align}
  (p-1) \gx^2 &= g^2~,\\
  \Rightarrow\quad \br(\phi \to xx) &= \frac{1}{p}~.
\end{align}
Then one finds
\begin{align}
  \ns + \De\ns &\propto \br + \De \br \non\\
  &= \frac{\gx^2 (1 + \De \gx/\gx)^2}
      {\gx^2 (1 + \De \gx/\gx)^2 + (p-1) \gx^2} \non \\
  &= \frac{1}{p} \LP 1 + 2 \frac{\De\gx}{\gx} - 2 \frac{\De\gx}{\gx\,p} \RP \\
\Rightarrow\quad \De\br &= \frac{1}{p}
      \LP 2 \frac{\De\gx}{\gx} - 2 \frac{\De\gx}{\gx\,p} \RP~, \\
\Rightarrow\quad \frac{\De\ns}{\ns} &= \frac{\De\br}{\br}      
        = 2 \frac{\De\gx}{\gx} - 2\,\frac{\De\gx}{\gx\,p}\non\\
  &= 2 \frac{\De\gx}{\gx} \LP 1 - \frac{1}{p} \RP~. 
\end{align}
  
\item[(ii)] The coupling is determined via the production cross
    section. This is the case for $\gZ$.

Then one can assume (where as above the $\br(Z \to \ell^+\ell^-)$
cancels out)
\begin{align}
  \ns &\propto \sig(e^+e^- \to \phi Z)
       \times \br(Z \to e^+e^-,\mu^+\mu^-) \propto \gZ^2~,\\
  \ns + \De\ns  &\propto \LP \gZ + \De\gZ \RP^2~.\\
   \De\ns/\ns &\propto 2 \frac{\De\gZ}{\gZ}~.
\end{align}
\end{itemize}


\subsubsection{Uncertainty in the Higgs couplings}

In the following we denote the SM Higgs boson as $h$, and the new Higgs
boson at $96 \gev$ as~$\phi$. 
For the SM Higgs-boson we have
\begin{itemize}
\item $\sig(e^+e^- \to Zh)$ from \refeq{sig-sm}
  and $\br(h \to xx)$ from \refta{tab:br-sm}, which gives us \Ns{h}.
\item $\LP \frac{\ns}{\nb} \RP_h$ from \refta{tab:signalbackground}.\\
  This allows us to evaluate $\LP \frac{\De\ns}{\ns} \RP_h$ via \refeq{eq:dns}.
\item $\LP \frac{\De\gx}{\gx} \RP_h$ from \refta{tab:gx-sm}.
\end{itemize}
For the new Higgs boson $\phi$ we have
\begin{itemize}
\item \Ns{\phi} from \refeq{eq:ns}.
\item For $\LP \frac{\ns}{\nb} \RP_\phi$ we assume
  $f_h/f_\phi = D$
  with $D = 2$ as starting/central point.\\
  This allows us to evaluate $\LP \frac{\De\ns}{\ns} \RP_\phi$ via
  \refeq{eq:dns}. Here it should be kept in mind that $D$ is apriori unknown.
  We use, as discussed above, $D = 3$ as a conservative value. 
\end{itemize}
Using the proportionality relations one can evaluate the coupling
precision in the two cases:
\begin{itemize}
  \item[(i)]
The coupling determined via the decay $\phi \to xx$. Here one finds, 
\begin{align}
  \frac{\LP \frac{\De \gx}{\gx} \RP_\phi}{\LP \frac{\De \gx}{\gx} \RP_h}
  &=
  \frac{\LP \frac{\De \ns}{\ns} \RP_\phi}{\LP \frac{\De \ns}{\ns} \RP_h}\times
  \frac{\LP 1 - \frac{1}{p_h} \RP}{\LP 1 - \frac{1}{p_\phi} \RP}
\end{align}
and can thus evaluate $\LP \frac{\De \gx}{\gx} \RP_\phi$ using
\begin{align}
   \frac{\LP \frac{\De \ns}{\ns} \RP_\phi}{\LP \frac{\De \ns}{\ns} \RP_h}
    &\times \frac{\LP 1 - \frac{1}{p_h} \RP}{\LP 1 - \frac{1}{p_\phi} \RP}
  = \frac{\LP \frac{\sqrt{1 + 1/f_\phi}}{\sqrt{N_{S,\phi}}} \RP}
       {\LP \frac{\sqrt{1 + 1/f_h}}{\sqrt{N_{S,h}}} \RP} \times
  \frac{\LP 1 - \frac{1}{p_h} \RP}{\LP 1 - \frac{1}{p_\phi} \RP} \\[.2em]
  &= \frac{\sqrt{1 + D/f_h}}{\sqrt{1 + 1/f_h}} \times
     \frac{\sqrt{N_{S,h}}}{\sqrt{N_{S,\phi}}} \times
     \frac{(1 - \br(h \to xx))}{(1 - \br(\phi \to xx))}\\[.2em]
  &= \sqrt{\frac{D + f_h}{1 + f_h}} \times
     \sqrt{\frac{\sig(e^+e^- \to Zh)}{\sig(e^+e^- \to Z\phi)}} \times
     \sqrt{\frac{\br(h \to xx)}{\br(\phi \to xx)}} \times
     \frac{(1 - \br(h \to xx))}{(1 - \br(\phi \to xx))}~.
\end{align}

\item[(ii)] The coupling is determined via the production cross
  section, i.e.\ \gZ. Here we find
  \begin{align}
  \frac{\LP \frac{\De \gZ}{\gZ} \RP_\phi}{\LP \frac{\De \gZ}{\gZ} \RP_h}
  &=
  \frac{\LP \frac{\De \ns}{\ns} \RP_\phi}{\LP \frac{\De \ns}{\ns} \RP_h}
\end{align}
and can thus evaluate $\LP \frac{\De \gZ}{\gZ} \RP_\phi$ using
\begin{align}
  \frac{\LP \frac{\De \ns}{\ns} \RP_\phi}{\LP \frac{\De \ns}{\ns} \RP_h}
  &= \frac{\sqrt{N_{S,h}}}{\sqrt{N_{S,\phi}}}\\[.2em]
  &= \sqrt{\frac{\sig(e^+e^- \to Zh)}{\sig(e^+e^- \to Z\phi)}}~.
\end{align}

\end{itemize}
\end{appendices}

\newpage

\bibliographystyle{unsrt}
\bibliography{reference}

\end{document}